\documentclass[a4paper,11pt]{article}
\usepackage[utf8]{inputenc}
\usepackage{jheppub,float}
\usepackage[normalem]{ulem}
\usepackage{xcolor}
\usepackage{amsmath}
\usepackage{framed}
\usepackage{varioref}
\colorlet{shadecolor}{lightgray}
\usepackage{hyperref}
\usepackage{mathrsfs}
\usepackage{parskip}
\labelformat{equation}{Eq.~(#1)}

\author[a]{Nabamita Banerjee,}
\author[b,c]{Karan Fernandes,}
\author[d]{and Arpita Mitra.}

\affiliation[a]{Indian Institute of Science Education \& Research Bhopal,\\
	Bhopal Bypass Road, Bhauri, Bhopal 420 066,\\ Madhya Pradesh, India.}
\affiliation[b]{Department of Physics, National Taiwan University, Taipei 10617, Taiwan.}
\affiliation[c]{Center of Astronomy and Gravitation, National Taiwan Normal University, Taipei 11677, Taiwan}
\affiliation[d]{{Physics Division, National Center for Theoretical Sciences, Taipei 10617, Taiwan}}
\emailAdd{nabamita@iiserb.ac.in, karanfernandes86@gmail.com, arpitamitra89@gmil.com}
\title{$1/L^2$ corrected soft photon theorem from a CFT$_3$ Ward identity}
\abstract{Classical soft theorems applied to probe scattering processes on AdS$_4$ spacetimes predict the existence of $1/L^2$ corrections to the soft photon and soft graviton factors of asymptotically flat spacetimes. In this paper, we establish that the $1/L^2$ corrected soft photon theorem can be derived from a large $N$ CFT$_3$ Ward identity. We derive a perturbed soft photon mode operator on a flat spacetime patch in global AdS$_4$ in terms of an integrated expression of the boundary CFT current. Using the same in the CFT$_3$ Ward identity, we recover the $1/L^2$ corrected soft photon theorem derived from classical soft theorems.} 
\begin{document} 
\maketitle
\flushbottom

\section{Introduction}
Soft theorems relate amplitudes with soft emission to the amplitude without the soft particles through a soft factor \cite{Weinberg:1964ew, Weinberg:1965nx}. In the case of the single soft photon theorem, a scattering amplitude ${\cal{A}}_{n+1}$ involving $m$ incoming hard particles, $n-m$ outgoing hard particles and a single soft external photon can be expressed as
\begin{align}
{\cal{A}}_{n+1} &= \langle p_{m+1}\, \cdots\, p_n \vert a_{\alpha}(q) \mathcal{S} \vert p_1 \, \cdots \,p_m  \rangle\notag\\
& = S^{\text{photon}} \langle p_{m+1}\,, \cdots\,, p_n \vert \mathcal{S} \vert p_1 \,, \cdots \,,p_m  \rangle = S^{\text{photon}} {\cal{A}}_{n} \,.
\label{spt}
\end{align}
In \ref{spt}, $\mathcal{S}$ is the $S$-matrix that relates incoming and outgoing states,  $p_1\,, \cdots\,, p_n$ are the momenta of the hard particles with charges $Q_1\,, \cdots Q_n$, $a_{\alpha}(q)$ is the soft photon creation operator in the outgoing state with momentum $q$ and polarization $\epsilon_{\alpha}$ and $S^{\text{photon}}$ is the soft photon factor which is a function of all the charges, momenta and the soft photon polarization. The soft factor admits an expansion in soft momentum $q$, with the leading pole contribution being the Weinberg soft photon factor \cite{Weinberg:1965nx}
\begin{equation}
S^{\text{photon}}_{(0)} = \sum_{i=m+1}^n Q_i \frac{p_i.\epsilon}{p.q} - \sum_{i=1}^m Q_i \frac{p_i.\epsilon}{p.q} \,.
\end{equation}
Remarkably, soft theorems have also been recently realized as a symmetry of the $S$-matrix along future and past null infinity on asymptotically flat spacetimes \cite{Strominger:2013jfa, Kapec:2014zla, He:2014cra, Cachazo:2014fwa, Campiglia:2015yka, Strominger:2017zoo}. More specifically, there exist soft charges that generate large gauge transformations of asymptotic massless fields, with the $S$-matrix satisfying the corresponding large gauge Ward identity \cite{Lysov:2014csa, Schwab:2014xua, Campiglia:2014yka, Casali:2014xpa, Conde:2016csj, Chakrabarti:2017zmh, Chakrabarti:2017ltl, Laddha:2017vfh, AtulBhatkar:2018kfi, Miller:2021hty}. Thus soft theorems are equivalent to large gauge Ward identities, with the intepretation of soft particles as Goldstone modes. This equivalence is part of a larger web of relations known as the `infrared triangle' for interacting theories with massless fields on asymptotically flat spacetimes~\cite{Strominger:2017zoo}.

It is currently unknown if similar infrared structures of scattering processes are present on non-asymptotically flat spacetimes, particularly those with a cosmological constant. While certain generalizations of BMS symmetries on spacetimes with a cosmological constant are known to result from modified boundary conditions~\cite{Compere:2019bua,Compere:2020lrt,Fiorucci:2020xto}, their relevance in scattering processes remain obscure. In large part, this is due to the absence of well defined scattering amplitudes on these spacetimes. However, the flat spacetime $S$-matrix can be recovered from the large radius limit of AdS correlation functions~\cite{Polchinski:1999ry, Giddings:1999jq, Gary:2009ae,Gary:2009mi,Penedones:2010ue,  Fitzpatrick:2011jn, Fitzpatrick:2011ia}. The scattering in this context is in a small locally asymptotically flat region within a larger AdS spacetime. In addition, some infrared properties of asymptotically flat spacetime $S$-matrices have been recovered in this large AdS radius limit, including the derivation of BMS symmetries~\cite{Hijano:2019qmi} and the soft photon theorem from CFT Ward identities~\cite{Hijano:2020szl}. Let us summarize how the correspondence works :
\begin{itemize}
\item[] Applying the bulk reconstruction method, one first constructs bulk AdS operators from the boundary CFT operators. Subsequently, the large AdS radius limit of these suitably constructed operators provide the corresponding flat spacetime creation and annihilation operators.
\end{itemize}
In particular, as we will see in detail in the main draft, for boundary vector current operators described in global coordinates one gets the photon creation and annihilation modes defined on a flat patch at the center of the AdS spacetime. 
Hence aspects of the asymptotically flat spacetime infrared triangle can in principle be derived from CFT correlation functions on asymptotically AdS spacetimes.

Given these results, it is natural to consider subleading large AdS radius corrections that could incorporate the effect of small cosmological constant to known flat spacetime results. It is not immediately clear if there exist $1/L$ corrections to the $S$-matrix or soft factors recovered in $L \to \infty$ limit of the spacetime, where $L$ is the AdS radius. An approach to derive classical contributions in soft factors without recourse to an $S$-matrix comes from classical soft theorems~\cite{Laddha:2018rle,Laddha:2018myi,Laddha:2018vbn,Laddha:2019yaj,Saha:2019tub,Fernandes:2020tsq}. These theorems provide universal soft factor contributions from classical scattering processes whenever the soft radiation has a wavelength much larger than the impact parameter and total energy far less than that of the scatterer. Classical soft theorems have been used to derive the soft graviton and soft photon factors on asymptotically flat spacetimes \cite{Laddha:2018myi}. Using a probe scattering process in the small cosmological constant limit, this approach was also used to derive universal soft graviton \cite{Banerjee:2020dww} and soft photon \cite{Banerjee:2021llh} factor contributions on asymptotically AdS$_4$ spacetimes up to the first subleading order.  At both leading ($\omega^{-1}$) and subleading ($\ln \omega^{-1}$) orders in frequency, there exist AdS radius specific corrections to the known universal asymptotically flat spacetime results.

The AdS radius dependent corrections are a consequence of a double scaling limit required on spacetimes with a cosmological constant. The spectrum of massless fields on AdS spacetimes is actually discrete and a typical $\omega \to 0$ limit does not exist. As further elaborated in \cite{Banerjee:2020dww, Banerjee:2021llh}, we can rather consider a double scaling limit wherein $\omega \to 0$ as $L \to \infty$ while leaving $\omega L = \gamma$ constant and large. In retaining all $1/L^2$ corrections, the radiation is seen to involve $1/L^2$ corrections while the massive probe particle asymptotic trajectory is corrected at the next subleading order ($1/L^4$). This is a consequence of the AdS$_4$ potential considered perturbatively about flat spacetime up to $1/L^2$ order, which affects the trajectories of massless particles while preserving the flat spacetime geodesics of massive particles. Hence the double scaling limit applied in conjunction with the classical soft theorems provide $1/\gamma^2$ corrections to soft factors on asymptotically AdS$_4$ spacetimes.  The leading soft photon factor was also shown to equivalent to a perturbed large gauge Ward identity on the asymptotically flat spacetime patch embedded in AdS spacetimes \cite{ Banerjee:2021llh}. It should also be noted that such corrections can't be replicated by higher curvature contributions to General Relativity on asymptotically flat spacetimes, as the corresponding soft factors in this case appear at subleading order in soft frequency. Hence $1/\gamma^2$ corrected soft factors provide evidence for asymptotic interactions that distinguish soft factorization of amplitudes between asymptotically flat spacetimes and those embedded in a larger ambient spacetime.

In this paper, we derive the leading $1/\gamma^2$ corrected soft photon factor on asymptotically AdS$_4$ spacetimes from boundary CFT Ward identities. We follow the approach in \cite{Hijano:2020szl}, wherein the asymptotically flat spacetime soft photon factor was derived from the $L \to \infty$ limit of a conserved $U(1)$ CFT in the large $N$ limit. Within this Lorentzian approach, the HKLL (Hamilton, Kabat, Lifschytz, and Lowe) bulk reconstruction~\cite{Hamilton:2006az} provides creation and annihilation soft photon modes in the flat spacetime scattering process from the dual $U(1)$ current smeared around a small window of global time. We extend this reconstruction to $1/L^2$ contributions, which through the double scaling limit, provide $1/\gamma^2$ corrections to flat spacetime results. The general expression for the corrected soft photon modes involve an integration over angles on the flat spacetime patch and CFT boundary.  We show that a leading contribution from this integral precisely agrees with the $1/\gamma^2$ corrected soft photon factor from classical soft theorems. 

The organization of our paper is as follows. In the next section, we review essential features and derivation of the Weinberg soft photon theorem from a large $N$ CFT$_3$ Ward identity on AdS$_4$ spacetimes following \cite{Hijano:2020szl}. In Sec. 3, we then consider $1/L^2$ corrections to the flat spacetime limit of the CFT Ward identity. We first review the result for $1/\gamma^2$ corrections to the flat spacetime soft factors coming from classical soft theorems. We then proceed to derive the corrected soft photon theorem from the CFT$_3$ Ward identity. We conclude the paper with some interesting open questions.

\section{Soft photon theorem from CFT Ward identities} \label{sec2}
In this section, we will review the derivation of Weinberg's soft photon theorem from a CFT$_3$ Ward identity at the boundary of AdS$_4$ spacetimes closely following \cite{Hijano:2020szl}. Experts familier with notations and other relevant details may skip this part.We first address certain preliminaries needed for this derivation before turning to the result from the Ward identity. The AdS$_4$ spacetime metric in global coordinates is
\begin{equation}
ds^2 = \frac{L^2}{\cos^2(\rho)}\left[- d\tau^2 + d \rho^2 + \sin^2(\rho) d\Omega_2^2\right]\,,
\label{ads.met2}
\end{equation}
where the $2$-sphere metric $d\Omega_2^2$ will be described using complex stereographic coordinates $\{z\,,\bar{z}\}$. The metric for the asymptotically flat spacetime patch follows from \ref{ads.met2} by defining
\begin{equation}
\frac{r}{L} = \tan (\rho) \qquad \;; \qquad  \frac{t}{L} = \tau \,, \label{t.tran}
\end{equation}
and taking $L \to \infty$. This patch is centrally located at global time $\tau = 0$.

The Lorentzian analysis in~\cite{Hijano:2020szl} is based on relating Fock states on the asymptotic boundary of the flat spacetime patch with CFT states on the boundary of the AdS spacetime. This is achieved through global Cauchy slices that foliate the spacetime, whose $L \to \infty$ limit recovers the asymptotic slices of the flat spacetime patch. In particular, the scattering process is confined to the flat patch in a region of $\epsilon \sim \mathcal{O}(L)^{-1}$ around global time $\tau = 0$. Beyond this region, the bulk fields are asymptotically free and their reconstruction from boundary operators can be carried out using the HKLL reconstruction. The outgoing (ingoing) states can be defined on Cauchy slices for $\tau \in \{\epsilon\,, \pi - \epsilon\}$ ($\tau \in \{-\pi + \epsilon\,, -\epsilon \}$) denoted by $\Sigma^+$ ($\Sigma^-$). 

We consider bulk gauge field operators $\hat{\mathcal{A}}_{\mu} (\rho \,,\tau\,,z\,,\bar{z})$ defined on early and late time slices, with the general boundary behaviour

\begin{equation}
\hat{\mathcal{A}}_{\mu} (\rho\,,x') \xrightarrow[\rho \to \frac{\pi}{2}]{} (\cos \rho)^1 \hat{\alpha}_{\mu}(x') +    (\cos \rho)^0 \hat{\beta}_{\mu}(x')\,,
\end{equation}

where $x' = \{\tau'\,,z'\,,\bar{z}'\}$ denotes global boundary coordinates, while $\hat{\alpha}_{\mu}(x')$ and $\hat{\beta}_{\mu}(x')$ are primary operators of the CFT that are sources. In the following, we fix $\hat{\beta}_{\mu}(x')$ to be a non-dynamical boundary gauge field that couples to a $U(1)$ conserved current $j_{\mu}(x')$ of conformal dimension $\Delta = 2$. This effectively chooses $\hat{\beta}_{\mu}(x')=0$ in the bulk and $\hat{\alpha}_{\mu}(x') = j_{\mu}(x')$. With these assumptions, the boundary limit of the bulk gauge field is simply

\begin{equation}
\hat{\mathcal{A}}_{\mu} (\rho\,,x') \xrightarrow[\rho \to \frac{\pi}{2}]{} (\cos \rho) j_{\mu}(x')\,,
\label{A.bnd}
\end{equation}

The choice in \ref{A.bnd} corresponds to `magnetic boundary conditions' and will provide us with Weinberg's soft photon theorem in the absence of magnetic charges. We further assume the absence of Coulombic fields, with the conserved current dual to radiative modes. Hence $\mu=z\,, \bar{z}$ provide the only non-vanishing current components.

In the $L \to \infty$ limit, bulk radiative fields must also satisfy the mode expansion on constant time slices of asymptotically flat spacetimes. To this end, we adopt the conventions of \cite{Strominger:2017zoo} and write the mode expansion as,
\begin{equation}
\hat{\mathcal{A}}_{\mu} (y) = \int \frac{d^3 \vec{q}}{(2 \pi)^3}\frac{1}{2 \omega_q} \sum_{\lambda=\pm} \left[ \varepsilon^{(\lambda)*}_{\mu} \hat{a}_{\vec{q}}^{(\lambda)} e^{i q y} + \varepsilon^{(\lambda)}_{\mu} \hat{a}_{\vec{q}}^{(\lambda) \dagger} e^{-i q y}\right] \,,
\label{A.flat}
\end{equation}

where $y = \{t\,, \vec{y}\} =  \{t\,, r\,,z\,,\bar{z}\}$ are flat spacetime coordinates, $q = \{q^0 \,, \vec{q}\}$ is the 4-momentum of the radiative fields satisfying $q^2 = 0$ with frequency $\omega_q$, and $\varepsilon^{(\lambda)}_{\mu}$  are the polarization vectors normalized according to $\varepsilon^{(+)}_{\mu}\varepsilon^{(-) \mu} = 1$. The expression in \ref{A.flat} can be used to derive the creation and annihilation operators
\begin{align}
 \hat{a}_{\vec{q}}^{(\lambda)} &= \lim_{t\rightarrow\pm\infty} i \int d^3 \vec{y} \,  \varepsilon^{(\lambda) \mu}  e^{-i q\cdot y} \overleftrightarrow{\partial_0} \hat{{\cal A}}_{\mu}(y) \,, \label{an.fl} \\
\hat{a}_{\vec{q}}^{(\lambda) \dagger} &= \lim_{t\rightarrow\pm\infty} - i \int d^3 \vec{y} \,  \varepsilon^{(\lambda) * \mu}  e^{i q\cdot y} \overleftrightarrow{\partial_0} \hat{{\cal A}}_{\mu}(y) \,.\label{cr.fl}
\end{align}

In \ref{an.fl} and \ref{cr.fl} the outgoing modes follow from $t \to \infty$ while the ingoing modes are those from $t \to -\infty$. 
The creation and annihilation modes of the ingoing and outgoing states satisfy
\begin{equation}
\left[\hat{a}_{\vec{q}}^{(\lambda)} \,,  \hat{a}_{\vec{q}\,'}^{(\lambda')}\right] = \delta^{\lambda \lambda'} (2 \pi)^3 2\omega_{q} \delta^{(3)}(\vec{q} - \vec{q}\,')
\label{com.rel}
\end{equation}

We can recover soft photon modes from \ref{an.fl} and \ref{cr.fl} in the $\omega_q \to 0$ limit. The outgoing positive helicity soft photon mode will result from \ref{an.fl}, for which we have the following polarization vector and plane wave expressions

\begin{align}
\varepsilon^{(+) z}  =  \frac{1 + z \bar{z}}{\sqrt{2} r}, \quad 
 e^{-i q\cdot y}  = e^{i \omega_{\vec{q}} t} e^{-i \vec{p}.\vec{r}} = e^{i \omega_{\vec{q}} t} 4\pi \sum_{l',m'} (-i)^{l'} j_{l'}(r \omega_{\vec{q}}) Y_{l'm'}(\Omega) Y^*_{l'm'}(\Omega_{q})\,,
\label{pl.exp1}
\end{align}

In order to derive flat spacetime soft modes from a CFT, we must also relate the gauge field appearing in \ref{an.fl} and \ref{cr.fl} with the CFT current at the boundary. This can be facilitated by using the HKLL prescription \cite{Hamilton:2006az}, with the reconstruction of bulk gauge fields in global coordinates that satisfy~\ref{A.bnd} taking the form

\begin{align}
\hat{{\cal A}}_{\mu}(X) =& \int d^3 x' \left[K^V_{\mu} (X\,;x') \epsilon_{\tau'}^{a' b'} \nabla_{a'} j^+_{b'} + K^S_{\mu}(X \,;x') \gamma^{a' b'} \nabla_{a'} j^+_{b'} \right.  \notag\\
&  + \left. \left(K^V_{\mu}\right)^*(X \,;x') \epsilon_{\tau'}^{a' b'} \nabla_{a'} j^-_{b'} + \left(K^S_{\mu}\right)^*(X \,;x') \gamma^{a' b'} \nabla_{a'} j^-_{b'} \right]\,,
\label{kern.max}
\end{align}
where $x' = \{\tau'\,,z'\,,\bar{z}'\}$ are boundary coordinates while $X = \{\tau \,, \rho \,, z\,,\bar{z}\}$ is a bulk point in global AdS$_4$ coordinates, $\epsilon^{a' b' c'}$ and $\nabla_{a'}$ are respectively the Levi-Civita tensor and covariant derivatives on the boundary, and $j^{\pm}_{b'}$ represent current components at the boundary with the $\pm$ signs indicating positive and negative frequency solutions. The explicit form of the boundary integral is
\begin{equation} 
\int d^3 x' = \int\limits_{\mathcal{T}} d\tau' \int d \Omega'  
\end{equation}
with the domain of integration $\mathcal{T}$ in the $\tau'$ integral being $\{-\pi\,,0\}$ for ingoing states and $\{0\,,\pi\}$ for outgoing states. Lastly, $K^V_{\mu}$ and $K^S_{\mu}$ appearing in \ref{kern.max} are respectively the HKLL kernels for `vector' and `scalar' type components of the Maxwell field. For the purely radiative modes we have the components 

\begin{align}
K^V_{z} (X\,;x') &= \frac{1}{\pi} \sum_{\kappa\,, l\,, m} \mathcal{N}^V  Y^*_{lm}\left(\Omega'\right) \partial_{z}Y_{lm}\left(\Omega \right) \Xi_{\kappa l}(\rho\,,\tau\,,\tau')\Big\vert_{\Delta =2} \label{kv.1}\\
K^V_{\bar{z}} (X\,;x') &= - \frac{1}{\pi} \sum_{\kappa\,, l\,, m} \mathcal{N}^V  Y^*_{lm}\left(\Omega'\right) \partial_{\bar{z}}Y_{lm}\left(\Omega \right) \Xi_{\kappa l}(\rho\,,\tau\,,\tau')\Big\vert_{\Delta =2} \label{kv.2}\\
K^S_{z} (X\,;x') &=  \frac{1}{\pi} \sum_{\kappa\,, l\,, m} \mathcal{N}^S Y^*_{lm}\left(\Omega'\right) \partial_{z}Y_{lm}\left(\Omega \right) \Xi_{\kappa l}(\rho\,,\tau\,,\tau')\Big\vert_{\Delta =1} \label{ks.1}\\
K^S_{\bar{z}} (X\,;x') &=  \frac{1}{\pi} \sum_{\kappa\,, l\,, m} \mathcal{N}^S Y^*_{lm}\left(\Omega'\right) \partial_{\bar{z}}Y_{lm}\left(\Omega \right) \Xi_{\kappa l}(\rho\,,\tau\,,\tau')\Big\vert_{\Delta =1} \;, \label{ks.2}
\end{align}
with
\begin{equation}
\Xi_{\kappa l}(\rho\,,\tau\,,\tau') = e^{i \omega_{\kappa} (\tau - \tau')} \sin^{l+1} \rho \cos^{\Delta-1} \rho \,_2F_1\left(- \kappa \,, \kappa + \Delta + l \,, \Delta - \frac{1}{2}\Big\vert \cos^2 \rho \right)\,,
\label{xi.def}
\end{equation}
and $\kappa, l, m$ are positive integers. Our expressions follow from the free Maxwell field solutions \cite{Ishibashi:2004wx}, which we review in Appendix \ref{appa}. The scaling dimensions $\Delta = 2$ and $\Delta=1$ are those of the vector and scalar type solutions. The frequency modes of fields in AdS$_4$ are discrete and related to the scaling dimension in the above solutions by
\begin{equation}
\omega_{\kappa} =  2 \kappa + \Delta + l 
\label{freq.ads}
\end{equation}

The normalizations $\mathcal{N}^V$ and $\mathcal{N}^S$ appearing in \ref{kv.1} - \ref{ks.2} are 
\begin{equation}
\mathcal{N}^V =  -\frac{1}{4 l(l+1)} \qquad \;; \qquad \mathcal{N}^S =  -\frac{1}{4 l(l+1)} \frac{i}{\omega_{\kappa} \vert_{\Delta =1}}\,.
\label{norm}
\end{equation}
This choice is consistent with the normalization of the CFT current in~\cite{Hijano:2020szl} and provides canonically normalized creation and annihilation operators in the flat spacetime patch. We can now substitute the $L \to \infty$ limit of \ref{kern.max} in \ref{an.fl} and \ref{cr.fl} to find expressions for the flat spacetime annihilation and creation operators in terms of derivatives of the boundary current. The evaluation of the $L \to \infty$ limit involves substituting for $\rho$ and $\tau$ using \ref{t.tran}. Consistency with the flat spacetime mode solutions also require that the discrete frequency modes $\omega_{\kappa}$ in \ref{freq.ads} scale with $L$ in the flat spacetime limit. This can be achieved by requiring that modes in $L \to \infty$ limit are dominated by large values of $\kappa$, with $\omega_{\kappa} \approx \omega L$ and where $\omega$ is the continuous frequency of modes in flat spacetime.  In this way, the sum over $\kappa$ gets traded for an integral over $\omega$ in the expressions \ref{kv.1} - \ref{ks.2}. Explicitly we have

\begin{equation}
2 \kappa = \omega_{\kappa} - \Delta - l \,, \qquad \sum_{\kappa} \to \frac{1}{2}\int d\omega L
\label{freq.flat}
\end{equation}

This procedure leads to a solution $\hat{{\cal A}}_{\mu}(y)$, with $y$ the Minkowski coordinates, from \ref{kern.max}. The flat spacetime modes $\hat{a}_{\vec{q}}^{(\pm) \text{out}}$ create photons with positive ($+$) and negative ($-$) helicity in the outgoing state and respectively result from $\hat{{\cal A}}^{\text{out}}_{z}(y)$ and $\hat{{\cal A}}^{\text{out}}_{\bar{z}}(y)$ in our conventions for flat spacetime modes. Denoting the corresponding annihilation modes as $\hat{a}_{\vec{q}}^{\text{out}(\pm)}$, we find that \ref{an.fl} gives the result

\begin{align}
\hat{a}_{\vec{q}}^{\text{out}(-)}=&   \frac{1}{4\omega_q} \frac{1+z_q \bar{z}_q}{\sqrt{2}}  \int d^3x' \,  e^{i \omega_{\vec{q}} L\left({\pi\over 2}-\tau' \right)}
 \frac{1}{\bar{z}_q-\bar{z}'} D^{z'}j^-_{z'}(x')\,,
\notag\\
\hat{a}_{\vec{q}}^{\text{out}(+)}=&  \frac{1}{4\omega_q} \frac{1+z_q \bar{z}_q}{\sqrt{2}}  \int d^3x' \,  e^{i \omega_{\vec{q}} L\left({\pi\over 2}-\tau' \right)}
 \frac{1}{z_q-z'} D^{\bar{z}'}j^-_{\bar{z}'}(x')
\label{atoj.flat}
\end{align}
The $\omega_q$ frequency is defined below \ref{A.flat}.We get the result in \ref{atoj.flat} after integrating over the general flat spacetime frequency $\omega$. \footnote{There is a delta function for the frequency from integrating over the spherical Bessel functions that picks the frequency $\omega_q$.}\\
The expressions for creation modes in the outgoing states and all modes in the ingoing states can be similarly derived from the bulk gauge field solution. The association of flat spacetime modes with current operators at the boundary has also been identified for massless and massive scalar fields in~\cite{Hijano:2020szl}. For the outgoing modes in \ref{atoj.flat} we see that the dominant contribution of the phase in the large $L$ limit comes around $\tau = \frac{\pi}{2}$. More generally for ingoing and outgoing massless fields in the large $L$ limit, the dominant contribution comes from a $\mathcal{O}(L)^{-1}$ region around $\tau = \pm \frac{\pi}{2}$.  This provides a correspondence between a small window around $\tau = \pm \frac{\pi}{2}$, denoted as $\tilde{\mathcal{I}}^{\pm}$, and null infinity on the flat spacetime patch $\mathcal{I}^{\pm}$. In the case of massive fields in the large $L$ limit, the phase has complex saddles around $\tau = \pm \frac{\pi}{2} \pm i f(\omega_p, m)$ with $f$ a function of the massive particle energy $\omega_p$ and mass $m$. This indicates that $i^{\pm}$ of the flat spacetime patch can be associated at the AdS$_4$ boundary with Euclidean caps that are analytic continuations in the global time from $\pm \frac{\pi}{2}$. The mapping between asymptotic regions of the flat spacetime patch and that of the AdS$_4$ boundary is indicated in Fig. 1
\begin{figure}[h]
    \begin{center}
        \includegraphics[scale = 0.3]{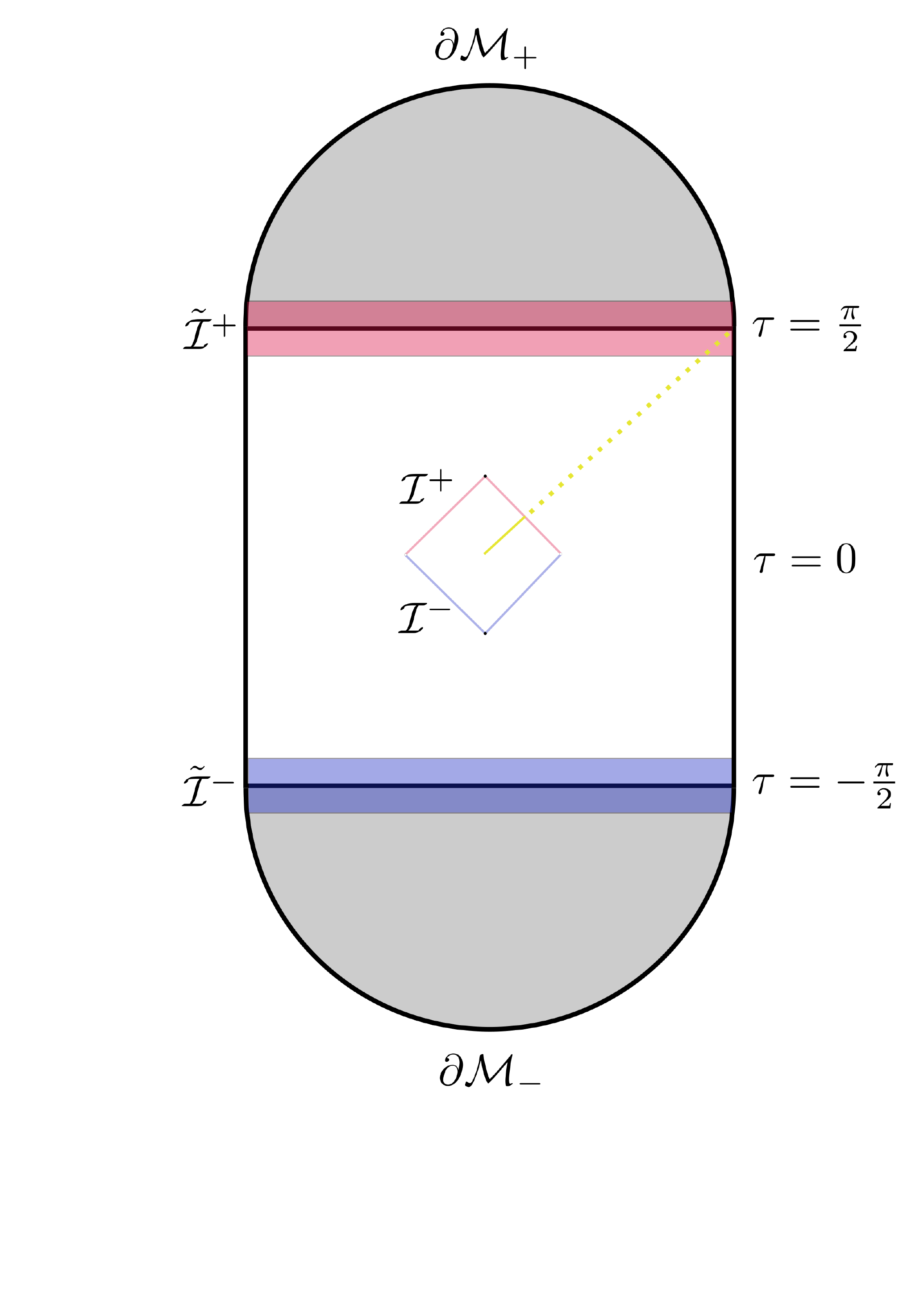}
    \end{center}
    \caption{The future (pink) and past (blue) null infinities of the central flat spacetime patch, $\mathcal{I}^+$ and $\mathcal{I}^-$ respectively, can be identified as the $L \to \infty$ limits of boundary regions $\tilde{\mathcal{I}}^{\pm}$ around $\tau' = \pm \frac{\pi}{2}$. Massless particles approaching the soft limit on the flat spacetime patch more closely approximate the $\tau' = \pm \frac{\pi}{2}$ surfaces. As an example, we have drawn a soft particle trajectory (yellow). The dashed line up to the boundary indicates the global trajectory outside the patch, but has no role in mapping the boundaries of AdS to those of the flat spacetime patch. Timelike infinities $i^{\pm}$ on the flat spacetime patch are identified with the Euclidean caps $\partial \mathcal{M}_{\pm}$, while spatial infinity $i_0$ is identified with $\tau \in \{-\frac{\pi}{2}\,,\frac{\pi}{2}\}$.}
\end{figure}

A feature of the Maxwell field modes which distinguish them from massless scalar field modes are the appearance of specific functions of the boundary angular coordinates $\{z'\,,\bar{z}'\}$. Denoting the parameter $\epsilon(\hat{x}')$ (with $\hat{x}'$ indicating dependence on angles) for the two helicity choices as 

\begin{equation}
\epsilon(\hat{x}') = \frac{1}{z_q - z'} \; (+\text{ve helicity}) \;, \qquad  \epsilon(\hat{x}') = \frac{1}{\bar{z}_q -\bar{z}'} \; (-\text{ve helicity}) \;,
\label{gp.def}
\end{equation}

we can express the $\omega_q \to 0$ limit of \ref{atoj.flat} as

\begin{align}
\lim_{\omega_q \to 0} \omega_q  \frac{\sqrt{2}}{1+z_q \bar{z}_q} \hat{a}_{\vec{q}}^{\text{out}(-)}=&   \frac{1}{4}  \int d^3 x' \epsilon(\hat{x}') D^{z'}j^-_{z'}(x')\,,
\notag\\
\lim_{\omega_q \to 0} \omega_q  \frac{\sqrt{2}}{1+z_q \bar{z}_q} \hat{a}_{\vec{q}}^{\text{out}(+)}=&  \frac{1}{4} \int d^3 x' \epsilon(\hat{x}') D^{\bar{z}'}j^-_{\bar{z}'}(x')
\label{atoj.soft}
\end{align}

The parameters in \ref{gp.def} are precisely those that are chosen in the large gauge Ward identity on asymptotically flat spacetimes to recover the Weinberg soft photon theorem \cite{Strominger:2017zoo}. We note that in taking the soft limit in \ref{atoj.soft}, the $\tau'$ dependent phase drops out. The soft limit hence has a boundary description on the $\tau = \pm \frac{\pi}{2}$ slices, providing a $2$ dimensional realization on the $3$ dimensional boundary. We also note that the outgoing positive (negative) helicity flat spacetime soft photon modes are mapped to $D^{\bar{z}} j_{\bar{z}}^-$ ($D^{z} j_{z}^-$) current derivatives on the AdS$_4$ boundary. 

With the above results, we can now derive Weinberg's soft photon theorem from the Ward identity of a large $N$ CFT with global $U(1)$ symmetry. The integrated expression for the Ward identity takes the form

\begin{equation}
\int d^3 x'\,\alpha(x')  \partial'_{\mu} \langle 0 \vert  T\{   j^{\mu}(x')  \Phi  \} \vert  0 \rangle  = \left(\sum_{i=1}^n Q_i \alpha (x'_i)- \sum_{j=1}^m   Q_j \alpha (x'_j) \right) \langle 0 \vert T\{ \Phi \}  \vert  0 \rangle
\, ,
\label{cftward}
\end{equation}

where $T\{\cdots \}$ refers to time ordering of the operators inside the parenthesis, $\Phi$ are a collection of CFT operators comprising of $n$ operators with charges $Q_i$ in the `ingoing' ($ \tau < 0 $) region and $m$ operators with charges $Q_j$ in `outgoing' ($\tau > 0$) region of the boundary, and $\alpha(x')$  an arbitrary parameter. Using the relationship between creation/annihilation flat spacetime modes with operators at the boundary, the correlation function $\langle 0 \vert T\{ \Phi \}  \vert  0 \rangle$ can be related with the $S$-matrix for a corresponding scattering process in the flat spacetime patch. 

The following choice for $\alpha(x')$

\begin{equation} 
\alpha(x')=\lim_{\rho\rightarrow{\pi \over 2}}\int d^2 \hat{x}'' \, {1\over 4\pi}{{\cos^2\rho-\cos^2\tau }\over{\left(   \sin\tau -\sin\rho\, \hat{x}'\cdot\hat{x}''  \right)^2}} \epsilon(\hat{x}'')\,,
\label{gpa.ads}
\end{equation}
recovers the soft theorem, as it has the desired property of $\alpha(x')\vert_{\tilde{\mathcal{I}}^{\pm}} = \epsilon(\hat{x}')$. Hence we recover the gauge parameters as in \ref{gp.def} which have no dependence on $\tau'$.  The left hand side of Weinberg's soft theorem, involving the insertion of the soft photon mode, follows from the left hand side of \ref{cftward}.

\begin{align}
\int d^3 x'\alpha(x')  \partial'_{\mu} \langle 0 \vert  T\{   j^{\mu}(x')  \Phi  \} \vert  0 \rangle = \int \limits_{\tilde{\mathcal{I}}^{\pm}} d^3 x' \epsilon(\hat{x}') \left[D^{z'}\langle 0 \vert  T\{   j_{z'}(x')  \Phi  \} \vert  0 \rangle + D^{\bar{z}'}\langle 0 \vert  T\{   j_{\bar{z}'}(x')  \Phi  \} \vert  0 \rangle \right]
\label{lhs}
\end{align}
The expression in \ref{lhs} can be directly associated with the soft photon insertion using \ref{atoj.soft} for the outgoing state. In summing over all positive and negative frequency contributions, only negative frequency terms contribute in the out-state. Hence the terms in \ref{atoj.soft} account for the insertion of a soft photon in the out-state of a given scattering process involving massless particles whose $S$-matrix results from $\langle 0 \vert T\{ \Phi \}  \vert  0 \rangle$. The procedure can be carried out for soft photons inserted in the in-state, with contributions in this case coming from positive frequency modes and thus creation operators. However, by invoking the equivalence of matrix elements involving in-state and out-state soft photons insertions by $CPT$ invariance, the contributions from the in-state can be readily related to the out-state soft photon insertions. 

The Weinberg soft photon theorem is recovered on considering \ref{gpa.ads} in the right hand side of \ref{cftward}. The derivation of the soft photon theorem  from a CFT Ward identity relied centrally on $\epsilon(\hat{x}')$ as in \ref{gp.def}, which was derived from the flat limit of HKLL reconstructed bulk gauge fields. We conclude this section with an observation of soft factors being encoded in \ref{atoj.flat} and consider only the outgoing positive helicity mode for simplicity. The flat spacetime parametrization for null particles can be applied to particles on the flat spacetime patch and AdS$_4$ boundary, as these only depend on angles. Assuming a `hard' massless particle with energy $E'$ and unit charge parametrized in terms of angular coordinates at the AdS$_4$ boundary, and a soft photon defined in terms of angular coordinates of the flat spacetime patch, we have

\begin{align}
p^{\mu} & = \frac{E'}{1 + z'\bar{z}'}\left(1+z'\bar{z}'\,, z'+ \bar{z}'\,, -i (z' - \bar{z}')\,, 1 - z'\bar{z}'\right) \,, \notag\\
q^{\mu} & = \frac{\omega_q}{1 + z_q\bar{z}_q}\left(1+z_q\bar{z}_q\,, z_q+ \bar{z}_q\,, -i (z_q - \bar{z}_q)\,, 1 - z_q\bar{z}_q\right) \notag\\
\epsilon^{\mu}_+ &= \frac{1}{\sqrt{2}} \left(\bar{z}_q\,, 1 \,, -i \,, -\bar{z}_q\right)\,,
\label{param.soft}
\end{align}

We then find that \ref{atoj.flat} is equivalent to the following expression

\begin{equation}
\hat{a}_{\vec{q}}^{\text{out}(+)} (\omega_q \hat{q}) = \frac{1}{4} \int d^3x' \,  e^{i \omega_{\vec{q}} L\left({\pi\over 2}-\tau' \right)}
 \frac{p.\epsilon_+}{p.q} \, D^{\bar{z}'}j^-_{\bar{z}'}(x') \,,
 \label{fl.gpsf}
\end{equation}

which involves the soft factor $ \frac{p.\epsilon_+}{p.q}$ for a positive helicity soft photon.  Similar expressions can be found for all other incoming and outgoing modes. This establishes that flat spacetime gauge field modes derived in terms of the boundary current contain information on the soft photon factor in soft theorems and the equivalent gauge parameter needed to derive the corresponding large gauge Ward identity.

\section{$1/L^2$ corrections to the soft photon theorem from CFT ward identities} \label{sec3}

We will now address $1/L^2$ corrections of the soft photon theorem on AdS$_4$ spacetimes. In the following subsection, we first briefly recall the soft factor correction derived previously using classical soft theorems on AdS$_4$ black hole spacetimes \cite{Banerjee:2020dww, Banerjee:2021llh}.  We will then proceed to generalize the above bulk reconstruction analysis up to $1/L^2$ corrections. The resulting expression for a perturbed soft photon mode in terms of a current can also be substituted in the Ward identity. In the last subsection, we establish that the corrected soft photon theorem resulting from the CFT$_3$ Ward identity agrees with the classical soft theorem result after expanding about a leading saddle.

\subsection{$1/L^2$ corrected soft photon theorem from classical soft theorems}

The formal derivation of soft factorization in scattering processes on asymptotically AdS spacetimes is obstructed by the absence of a globally defined $S$-matrix. This motivated our derivation of soft factors using classical soft theorems. These theorems state that the classical limit of soft photon and graviton factors may be derived from the zero frequency limit of certain classical scattering processes. More significantly, the primary requirement is that of gauge invariant observables in the case of electromagnetically mediated scattering and diffeomorphism invariance in gravitational scattering, without specifc reference to the background geometry. This allows for the derivation of soft factors on spacetimes with a cosmological constant \cite{Banerjee:2020dww, Banerjee:2021llh, AtulBhatkar:2021sdr}. The classical scattering processes are broadly constrained to be such that the energy of the scatterer does not significantly change during the scattering process ($\Delta E_{\text{scatterer}}<<1$) and the wavelength of the emitted radiation should be greater than the large impact parameter ($\lambda_{\text{radiation}} >> b$). 

Among the scattering processes that satisfy these criteria are probe scattering processes on curved spacetimes. We accordingly considered the scattering of a probe particle on asymptotically AdS black hole spacetimes in \cite{Banerjee:2020dww, Banerjee:2021llh}.  The existence of a largest length scale in the classical scattering, namely the AdS radius $L$, introduces two additional requirements in applying classical soft theorems. The first concerns the radial distance of the probe from the scatterer, which we denote by $r$. With the black hole radius $GM$, we require the large impact parameter requirement to be modified to $GM<<r<<L$ . Hence the classical process is confined to a region deep in the bulk of asymptotically AdS spacetimes. The second requirement comes in the derivation of the soft limit. Since $r$ cannot take on asymptotically large values, the scattering process takes place within a finite interval of time. In addition, the frequency of massless fields on AdS spacetimes is discrete, formally preventing a zero frequency limit. We hence implement a double scaling limit, wherein $\omega \to 0$ as $L \to \infty$, while keeping $\omega L = \gamma$ constant and large. \footnote{As compared to the last section,  $\gamma = \omega L \approx \omega_{\kappa}$, but their context is different.} 

We further note that in the classical soft photon theorem derivation, we never encountered the discrete frequency of AdS. The derivation is on an asymptotically flat spacetime perturbed by small cosmological constant corrections. Hence $\omega$ and $L$ were given  and we defined a suitable double scaling limit where their product is a large constant. \footnote{In the AdS derivation, the discrete mode can recover continuous flat spacetime modes in the large $L$ limit by double scaling. As noted in the previous section, here we have $\omega_{\kappa}$ and $L$ given, while $\omega$ is defined to be the continuous flat spacetime frequency.}

With these assumptions, the equations for the radiative fields were derived retaining all $1/L^2$ corrections. The double scaling limit applied to the radiative fields identified $1/\gamma^2$ corrections to soft factors on asymptotically flat spacetimes. More specifically, the leading ($\omega^{-1}$) and subleading ($\ln \omega^{-1}$) soft photon and soft graviton factors were derived, each with their respective $\frac{1}{\gamma^2}$ corrections. Furthermore, the $1/L^2$ corrections of the probe particle trajectory, while present, lead to contributions at subleading order in frequency. This suggests that $1/\gamma^2$ corrected soft factors would be those for an $S$-matrix on an asymptotically flat spacetime patch within a global AdS spacetime.

In the following, we restrict ourselves to the leading soft photon factor. The inferred form of the leading soft factor for a general process involving $n$ hard particles with momenta $p_{(a)}^{\mu}$ and charges $Q_{(a)}$, and a single soft photon with momentum $q^{\mu}$ and polarization $\epsilon_{\mu}$ takes the form
\begin{align}
     S^{(0)}_{\text{em}} &= S^{(0);\text{f}}_{\text{em}} + S^{(0);\text{L}}_{\text{em}} \notag\\
    S^{(0);\text{f}}_{\text{em}} &= \sum_{a=1}^n  Q_{(a)} \eta_{(a)} \frac{\epsilon_{\mu} p^{\mu}_{(a)}}{p_{(a)}.q} \,, \label{fl.spf}\\ 
    S^{(0);\text{L}}_{\text{em}} &= \frac{\omega^2}{4 \gamma^2} \sum_{a=1}^n Q_{(a)} \eta_{(a)} \frac{\epsilon_{\mu} p^{\mu}_{(a)}}{p_{(a)}.q} \frac{\vec{p}^2_{(a)}}{\big(p_{(a)}.q\big)^2} \,,
\label{cst.spf}
\end{align}
where $\eta_{(a)} = 1(-1)$ for outgoing (ingoing) hard particles. All indices in the above expressions are contracted with the flat spacetime metric. 

\color{black}

As the leading soft factor is universal and holds beyond tree level, we can consider the above expression at the level of the soft photon theorem in a flat spacetime scattering process
 \begin{align}
 \lim_{\omega \rightarrow 0} \omega &\left(\langle \text{out}\vert \hat{a}^{\text{out}(+)}(\omega \hat{x})\mathcal{S} \vert \text{in} \rangle + n_L\langle \text{out}\vert \hat{a}^{\text{out; L}\,(+)}(\omega \hat{x})\mathcal{S} \vert \text{in} \rangle\right) = \left(S^{(0);\text{f}}_{\text{em}} + S^{(0);\text{L}}_{\text{em}}\right) \langle \text{out}\vert \mathcal{S} \vert \text{in} \rangle
 \label{spt.corr}
 \end{align}
with $\mathcal{S}$ the $S$-matrix of the scattering process and $n_L$ is an overall constant. The operators $\hat{a}^{\text{out}(+)}$ and $\hat{a}^{\text{out; L}\,(+)}$ are those for two positive helicity soft photon modes responsible for the corresponding soft factors  $S^{(0);\text{f}}_{\text{em}}$ and $S^{(0);\text{L}}_{\text{em}}$. Due to the absence of $1/L^2$ corrections of the probe particle trajectory in the derivation using classical soft theorems, we assume the soft factors are infrared divergent contributions to the uncorrected $S$-matrix on the asymptotically flat spacetime patch. This implies that $\hat{a}^{\text{out}(+)}$ is the usual soft photon mode leading to the Weinberg soft factor, while $\hat{a}^{\text{out; L}\,(+)}$ can be interpreted as a perturbed mode that provides the $1/L^2$ corrected soft factor $S^{(0);\text{L}}_{\text{em}}$. This interpretation is supported by the equivalence of the $1/L^2$ corrected soft photon theorem with a perturbed large gauge Ward identity.

We will now evaluate \ref{spt.corr} for the massless scattering process of interest in our paper. Assuming the parametrization of hard particles and a single soft photon as in \ref{param.soft}, we find the following correction to the soft photon theorem
\begin{align}
& \lim_{\omega_q \rightarrow 0} \frac{\sqrt{2} \omega_q n_L}{\left(1 + z_q \bar{z}_q\right)}\langle \text{out}\vert \hat{a}_{\vec{q}}^{\text{out; L}\, (+)}(\omega_q \hat{x}) \mathcal{S} \vert \text{in} \rangle \notag\\
 & \qquad =  \frac{1}{16 \gamma ^2}\left[\ \sum_{k = \text{out}} \frac{\left(1 + z' \bar{z}'\right)^2 \left(1 + z_q \bar{z}_q\right)^2}{(\bar{z}_q-\bar{z}')^2 (z_q-z')^3}Q_{k}  - \sum_{k=\text{in}} \frac{\left(1 + z' \bar{z}'\right)^2 \left(1 + z_q \bar{z}_q\right)^2}{(\bar{z}_q -\bar{z}')^2 (z_q - z')^3}Q_k \right] \langle \text{out}\vert \mathcal{S} \vert \text{in} \rangle
\label{spt.cst}
\end{align}

The above result is the $1/L^2$ corrected soft photon theorem inferred from a purely classical scattering process in the bulk of AdS$_4$ spacetimes, up to an overall constant $n_L$ that cannot be fixed by classical soft theorems.

\subsection{$1/L^2$ corrected soft photon theorem from a CFT$_3$ Ward identity}

We will now consider the approach in Section \ref{sec2} to derive $1/L^2$ corrections to the known soft photon theorem for a $S$-matrix defined on the asymptotically flat spacetime patch in AdS$_4$. This implies that we do not consider $1/L^2$ corrections to the $L \to \infty$ limit of the global AdS$_4$ metric in \ref{ads.met2}, nor the time ordered collection of fields $\Phi$ appearing in the Ward identity \ref{cftward}. In this way, the `hard process' remains one of the $S$-matrix on an asymptotically flat spacetime patch.\footnote{We would technically find $\frac{r^3}{L^2}$ corrections to the flat spacetime metric on expanding \ref{t.tran} and retaining terms up to $1/L^2$. We expect that in the presence of such corrections, a consistent definition of an $S$-matrix with a soft factor will require the soft graviton theorem, which lies outside the scope of the present article.} However, with insights from the classical soft photon theorem, we expect that a scattering process on a flat spacetime patch embedded in an AdS$_4$ spacetime admits $1/L^2$ corrections to the soft photon theorem. We systematically derive the corrected soft photon mode by expanding the integrand of \ref{kern.max} up to $1/L^2$ corrections, assuming that the current remains fixed by the condition in \ref{A.bnd}. The substitution of the bulk gauge field up to $1/L^2$ corrections in \ref{an.fl} and \ref{cr.fl} then recovers soft photon modes as in \ref{atoj.soft} along with its perturbation responsible for $1/L^2$ corrections to the soft photon factor.

From \ref{kv.1} - \ref{ks.2}, we see that the $1/L^2$ corrections to the integrand in \ref{kern.max} can come from the function $\Xi_{\kappa l}(\rho\,,\tau\,,\tau')$ and possibly modified normalizations $\tilde{\mathcal{N}}^{V}$ and $\tilde{\mathcal{N}}^{S}$ in the kernels. We first address the corrections coming from $\Xi_{\kappa l}(\rho\,,\tau\,,\tau')$ as defined in \ref{xi.def}. As we continue to work in the nearly flat spacetime limit by considering $L$ large, the relation between $\tau$ and $t$ remains the same as in \ref{ads.met2}. We hence consider the same replacement of discrete frequencies $\omega_{\kappa}$ in AdS by $\omega L$ around large values of $\kappa$ (as in the $L \to \infty$ limit) \footnote{This approximation in the large $L$ limit is consistent with the double scaling limit $\omega \to 0$ as $L \to \infty$ with $\omega L = \gamma$ a large constant used in the derivation of $1/\gamma^2$ soft factor corrections from classical soft theorems.}. 

We substitute $\omega_{\kappa}$ from \ref{freq.flat} as well as $\tau$ and $\rho$ from \ref{ads.met2} in the \ref{xi.def}, and expand up to $1/L^2$ terms. The technical details behind this expansion are provided in Appendix \ref{appb}. The final result in the vector and scalar type expressions are
\begin{align}
&\Xi_{\kappa l}(\rho\,,\tau\,,\tau')\Big \vert_{\Delta = 2} =  - (\pm i)^{- l} e^{i \omega t} e^{-i \omega L \left( \tau' \mp \frac{\pi}{2}\right)} \frac{r}{L}\Bigg\{j_{l}(r \omega) \left( 1 + \frac{1}{2 \omega^2 L^2}\left(\frac{l(l+1)}{2} - \frac{(r \omega)^2}{3}\right)\right) \notag\\ 
&   \qquad \qquad \qquad \qquad  - \frac{1}{2 \omega^2 L^2} \sqrt{\frac{\pi}{2 r \omega}} \left( \frac{l (l+1)}{2} + (r \omega)^2 \right) \frac{2 r\omega}{3} J'_{l+\frac{1}{2}}(r \omega) \Bigg\} + \mathcal{O}\left( \frac{1}{\omega^3 L^3}\right)\,,\label{xi.d2}\\
& \Xi_{\kappa l}(\rho\,,\tau\,,\tau')\Big \vert_{\Delta = 1} = - (\pm i)^{- l + 1}  e^{i \omega t} e^{-i \omega L \left( \tau' \mp \frac{\pi}{2}\right)} (\omega L) \frac{r}{L} \Bigg\{j_{l}(r \omega) \left( 1 + \frac{1}{2 \omega^2 L^2}\left(-\frac{l(l+1)}{2} - \frac{(r \omega)^2}{3} \right)\right) \notag\\ 
&  \qquad \qquad \qquad \qquad  - \frac{1}{2 \omega^2 L^2} \sqrt{\frac{\pi}{2 r \omega}} \left( \frac{l (l+1)}{2} + (r \omega)^2 \right) \frac{2 r\omega}{3} J'_{l+\frac{1}{2}}(r \omega)\Bigg\} + \mathcal{O}\left( \frac{1}{\omega^3 L^3}\right) \,,\label{xi.d1}
\end{align}
where $J_{k}(v)$ are Bessel functions of the first kind of order $k$ and argument $v$, while $j_l(v)$ are spherical Bessel functions defined as
\begin{equation}
j_l(v) = \sqrt{\frac{\pi}{2 v}} J_{l+\frac{1}{2}}(v)\,.
\label{sphb.def}
\end{equation}
 The primes on Bessel functions in \ref{xi.d2} and \ref{xi.d1} denote derivatives with respect to the argument.
 
The need for $1/L^2$ corrections to the normalizations in \ref{norm} comes from requiring that perturbed soft modes of positive (negative) helicity continue being related to $D^{\bar{z}} j_{\bar{z}}^-$ ($D^{z} j_{z}^-$) current derivatives on the AdS$_4$ boundary, as discussed below \ref{atoj.soft}. If we continue to use the normalizations in \ref{norm}, we in fact get the opposite identification. As discussed in the previous section, the individual flat spacetime soft modes are further associated with the gauge parameter and soft factors of the same helicity in the boundary integrated expression over the derivatives of the current. We take this relationship between modes and current components to be a constraint respected under perturbations. Up to a common shift term proportional to $\frac{1}{\omega^2 L^2}$ in both $\mathcal{N}^V$ and $\mathcal{N}^S$, this restricts the possible modifications of the corrected normalizations $\tilde{\mathcal{N}}^V$ and $\tilde{\mathcal{N}}^S$ to be either of two possibilities 

\begin{align}
\tilde{\mathcal{N}}^V &= \mathcal{N}^V \;  \quad \;; \quad \tilde{\mathcal{N}}^S =  \mathcal{N}^S \left(1 + \frac{l(l+1)}{2 \omega^2 L^2}\right)\,, \label{norm.1}\\
\tilde{\mathcal{N}}^V &= \mathcal{N}^V \left(1 - \frac{l(l+1)}{2 \omega^2 L^2}\right) \;  \quad \;; \quad \tilde{\mathcal{N}}^S =  \mathcal{N}^S \,,
\label{norm.2}
\end{align}

The soft factor results we would get from these normalizations agree up to a sign. We choose \ref{norm.1} in the following. We stress that the modified normalization is not motivated to satisfy a known normalization or inner product relation. Such a criteria does not exist for the perturbed modes we seek to derive about flat spacetimes. Rather, we infer this correction purely from requiring the consistency between helicity components in bulk flat spacetime modes and boundary currents is respected to $1/L^2$ corrections.
 
We hence find the expressions
\begin{align}
\tilde{\mathcal{N}}^V &\Xi_{\kappa l}(\rho\,,\tau\,,\tau')\Big \vert_{\Delta = 2} =  \frac{(\pm i)^{- l}}{4 l(l+1)} e^{i \omega t} e^{-i \omega L \left( \tau' \mp \frac{\pi}{2}\right)} \frac{r}{L}\Bigg\{j_{l}(r \omega) \left( 1 + \frac{1}{2 \omega^2 L^2}\left(\frac{l(l+1)}{2} - \frac{(r \omega)^2}{3}\right)\right) \notag\\ 
&   \qquad \qquad \qquad  - \frac{1}{2 \omega^2 L^2} \sqrt{\frac{\pi}{2 r \omega}} \left( \frac{l (l+1)}{2} + (r \omega)^2 \right) \frac{2 r\omega}{3} J'_{l+\frac{1}{2}}(r \omega) \Bigg\} + \mathcal{O}\left( \frac{1}{\omega^3 L^3}\right)\,, \label{xin.d2}\\
\tilde{\mathcal{N}}^S& \Xi_{\kappa l}(\rho\,,\tau\,,\tau')\Big \vert_{\Delta = 1} = -  \frac{(\pm i)^{- l}}{4 l(l+1)}  e^{i \omega t} e^{-i \omega L \left( \tau' \mp \frac{\pi}{2}\right)} \frac{r}{L} \Bigg\{j_{l}(r \omega) \left( 1 + \frac{1}{2 \omega^2 L^2}\left(\frac{l(l+1)}{2} - \frac{(r \omega)^2}{3} \right)\right) \notag\\ 
&  \qquad \qquad \qquad  - \frac{1}{2 \omega^2 L^2} \sqrt{\frac{\pi}{2 r \omega}} \left( \frac{l (l+1)}{2} + (r \omega)^2 \right) \frac{2 r\omega}{3} J'_{l+\frac{1}{2}}(r \omega)\Bigg\} + \mathcal{O}\left( \frac{1}{\omega^3 L^3}\right) \,,\label{xin.d1}
\end{align}
which can be susbtituted in \ref{kv.1} - \ref{ks.2} to find any $1/L^2$ corrected bulk gauge field component in either the ingoing $\tau<0$ or outgoing $\tau>0$ states. In the following, we confine ourselves to the derivation of the perturbed mode that creates a positive helicity soft photon in the outgoing state. This mode is derived from the $\hat{{\cal A}}^{\text{out}}_{z}(y)$ expression in the large $L$ limit, that takes the form
\begin{align}
\hat{{\cal A}}^{\text{out}}_{z}(y) = \hat{{\cal A}}^{\text{out; f}}_{z}(y) + \hat{{\cal A}}^{\text{out;} L}_{z}(y) + \hat{{\cal A}}^{\text{out; sub}}_{z}(y)\,,
\label{az.full}
\end{align}

with $y$ the coordinates on the flat spacetime patch. The $\hat{{\cal A}}^{\text{out; f}}_{z}(y)$, $\hat{{\cal A}}^{\text{out; L}}_{z}(y)$ and $\hat{{\cal A}}^{\text{out; sub}}_{z}(y)$ respectively denote the flat spacetime, leading $1/L^2$ and subleading contributions, with expressions

\begin{align}
\hat{{\cal A}}^{\text{out; f}}_{z}(y) =& \frac{1}{4\pi} \int\limits_{0}^{\pi} d\tau' \int d\Omega' \int d\omega\, r \, j_{l}(r \omega) \notag\\
& \qquad \left[ \sum_{l,m} \frac{Y^*_{lm}\left(\Omega'\right)}{-l(l+1)} \partial_{z}Y_{lm}\left(\Omega \right) ( i)^{- l} e^{i \omega t} e^{-i \omega L \left( \tau' - \frac{\pi}{2}\right)} D^{\bar{z}'} j^+_{\bar{z}'} \right.\notag\\
&\left. \qquad \qquad \qquad + \sum_{l,m} \frac{Y_{lm}\left(\Omega'\right)}{-l(l+1)} \partial_{z}Y^*_{lm}\left(\Omega \right) (- i)^{- l} e^{-i \omega t} e^{i \omega L \left( \tau' - \frac{\pi}{2}\right)} D^{\bar{z}'} j^-_{\bar{z}'}\right]
\label{az.flat}
\end{align}

\begin{align}
\hat{{\cal A}}^{\text{out; L}}_{z}(y) =& \frac{1}{16 \pi} \int\limits_{0}^{\pi} d\tau' \int d\Omega' \int d\omega\,\frac{r}{\omega^2 L^2} \, j_{l}(r \omega) l(l+1) \notag\\
& \qquad \left[ \sum_{l,m} \frac{Y^*_{lm}\left(\Omega'\right)}{-l(l+1)} \partial_{z}Y_{lm}\left(\Omega \right) ( i)^{- l} e^{i \omega t} e^{-i \omega L \left( \tau' - \frac{\pi}{2}\right)} D^{\bar{z}'} j^+_{\bar{z}'} \right.\notag\\
&\left. \qquad \qquad \qquad + \sum_{l,m} \frac{Y_{lm}\left(\Omega'\right)}{-l(l+1)} \partial_{z}Y^*_{lm}\left(\Omega \right) (- i)^{- l} e^{-i \omega t} e^{i \omega L \left( \tau' - \frac{\pi}{2}\right)} D^{\bar{z}'} j^-_{\bar{z}'}\right]
\label{az.ads}
\end{align}

\begin{align}
\hat{{\cal A}}^{\text{out; sub}}_{z}(y) &= - \frac{1}{24 \pi} \int\limits_{0}^{\pi} d\tau' \int d\Omega' \int d\omega \frac{r}{\omega^2 L^2}\left[\left(\frac{l(l+1)}{2} + (r \omega)^2 \right) \sqrt{2 \pi r \omega} \, J'_{l+\frac{1}{2}}(r \omega) + (r \omega)^2 j_{l}(r \omega)\right] \notag\\
& \qquad \left[ \sum_{l,m} \frac{Y^*_{lm}\left(\Omega'\right)}{-l(l+1)} \partial_{z}Y_{lm}\left(\Omega \right) ( i)^{- l} e^{i \omega t} e^{-i \omega L \left( \tau' - \frac{\pi}{2}\right)} D^{\bar{z}'} j^+_{\bar{z}'} \right.\notag\\
&\left. \quad \qquad + \sum_{l,m} \frac{Y_{lm}\left(\Omega'\right)}{-l(l+1)} \partial_{z}Y^*_{lm}\left(\Omega \right) (- i)^{- l} e^{-i \omega t} e^{i \omega L \left( \tau' - \frac{\pi}{2}\right)} D^{\bar{z}'} j^-_{\bar{z}'}\right] + \mathcal{O}\left( \frac{1}{\omega^3 L^3}\right)
\label{az.sub}
\end{align}

On substituting \ref{az.full} in \ref{an.fl} we recover corresponding  outgoing positive helicity gauge field modes in a flat spacetime scattering process. The mode corresponding to the $\hat{{\cal A}}^{\text{out; f}}_{z}(y)$ contribution is the same as in \ref{atoj.flat} and provides the mode that creates an outgoing photon, whose soft limit \ref{atoj.soft} recovers the Weinberg soft photon theorem through the $U(1)$ CFT$_3$ Ward identity as reviewed in the previous section.

On replacing the bulk field contribution $\hat{{\cal A}}^{\text{out; L}}_{z}(y)$ of \ref{az.ads} in \ref{an.fl}, we find a perturbed mode in flat spacetime that we denote by $\hat{a}_{\vec{q}}^{\text{out; L}\, (+)}$. This mode is perturbative and it involves corrections in terms of the dimensionless parameter $1/\gamma^2 = 1/(\omega_q L)^2$. The derivation of this mode is given in Appendix \ref{appc} with the result 

\begin{equation}
\hat{a}_{\vec{q}}^{\text{out; L}\, (+)} = \frac{1+z_q \bar{z}_q}{ \sqrt{2}\omega_q}\frac{1}{32 \pi \gamma^2} \int\limits_{0}^{\pi} d\tau' \int d\Omega' \int d\Omega_{w} \left[\frac{\left(1+ z' \bar{z}'\right)^2\left(1+ z_w \bar{z}_w\right)^2}{\left(\bar{z}' - \bar{z}_{w}\right)^2 \left(z_q - z_w\right)^3}\right] \mathcal{D}^{\bar{z}'}j^{-}_{\bar{z}'} e^{i \omega_q L \left( \tau' - \frac{\pi}{2}\right)}
\label{atoj.L}
\end{equation}

In repeating the above procedure for other ingoing and outgoing bulk field modes, we can likewise find the corresponding perturbed creation and annihilation operators on the flat spacetime patch. For instance, from the expression of $\hat{\mathcal{A}}^{\text{out}}_{\bar{z}}(y)$, we can find the perturbed negative helicity outgoing mode 

\begin{equation}
\hat{a}_{\vec{q}}^{\text{out; L}\, (-)} = \frac{1+z_q \bar{z}_q}{ \sqrt{2}\omega_q}\frac{1}{32 \pi \gamma^2} \int\limits_{0}^{\pi} d\tau' \int d\Omega' \int d\Omega_{w} \left[\frac{\left(1+ z' \bar{z}'\right)^2\left(1+ z_w \bar{z}_w\right)^2}{\left(z' - z_{w}\right)^2 \left(\bar{z}_q - \bar{z}_w\right)^3}\right] \mathcal{D}^{z'}j^{-}_{z'} e^{i \omega_q L \left( \tau' - \frac{\pi}{2}\right)}
\label{atoj.Ln}
\end{equation}

Apart from the inclusion of an overall factor involving $1/\gamma^2$ in these corrected modes, we draw attention to the additional integral over intermediate angles $\{w \,,\bar{w}\}$ in \ref{atoj.L} that is absent in the flat spacetime result in \ref{atoj.flat}. The appearance of intermediate angular integrals will generically be a property to all higher powers in $1/L^2$, as these terms involve higher order derivatives of the spherical harmonics. Such terms can be expressed in terms of derivatives acting on products of Green's functions on the $2$-sphere, with additional angular integrals as in \ref{atoj.L}. The recovery of the $1/L^2$ corrected soft photon theorem in \ref{spt.cst} from \ref{atoj.L} will be considered in the following subsection.

Lastly, the bulk field contribution in \ref{az.sub} (apart from the $\mathcal{O}\left( \frac{1}{\omega^3 L^3}\right)$ terms ignored in our analysis) contain terms that are subleading in frequency. More specifically, they provide $1/\gamma^2$ corrected terms with higher order $\omega_q$ contributions to the leading $\omega_q^{-1}$ soft factor in \ref{atoj.L}. Hence the total contribution from \ref{az.sub} is subleading in frequency to the leading $1/L^2$ corrected soft photon theorem.

\subsection{Recovering the classical soft photon theorem result} \label{match}
In this section we recover the classical soft photon results.
The perturbed soft photon modes can be derived by taking the soft limit, namely $\omega_q \to 0$. Taking this limit in \ref{atoj.L} and \ref{atoj.Ln}, we find the following soft operator mode expressions in terms of the CFT$_3$ current

\begin{align}
\lim_{\omega_q \to 0} \omega_q  \frac{\sqrt{2}}{1+z_q \bar{z}_q} \hat{a}_{\vec{q}}^{\text{out; L}\,(+)}=&  \frac{1}{4} \int d^3 x' \epsilon^{\text{L}}(\hat{x}') D^{\bar{z}'}j^-_{\bar{z}'}(x')\notag\\
\lim_{\omega_q \to 0} \omega_q  \frac{\sqrt{2}}{1+z_q \bar{z}_q} \hat{a}_{\vec{q}}^{\text{out; L}\,(-)}=&  \frac{1}{4} \int d^3 x' \epsilon^{\text{L}}(\hat{x}') D^{z'}j^-_{z'}(x')
\end{align}

with the gauge parameter for the positive and negative helicity cases now defined as

\begin{align}
\epsilon^{\text{L}}(\hat{x}') &= \frac{1}{8 \pi \gamma^2}\int d\Omega_{w} \left[\frac{\left(1+ z' \bar{z}'\right)^2\left(1+ z_w \bar{z}_w\right)^2}{\left(\bar{z}' - \bar{z}_{w}\right)^2 \left(z_q - z_w\right)^3}\right] \qquad (+\text{ve helicity}) \;, \notag\\  \epsilon^{\text{L}}(\hat{x}') &= \frac{1}{8 \pi \gamma^2}\int d\Omega_{w} \left[\frac{\left(1+ z' \bar{z}'\right)^2\left(1+ z_w \bar{z}_w\right)^2}{\left(z' - z_{w}\right)^2 \left(\bar{z}_q - \bar{z}_w\right)^3}\right] \qquad (-\text{ve helicity}) \;.
\label{gpa.adsc}
\end{align}

The above expressions seem to indicate that to find the gauge parameters, we need to integrate over the angles. But that would not be correct to do when we want to recover the classical limit of it. This is due to the difference between the standard large $L$ limit in our present analysis and the choice of isotropic coordinates used in the derivation of the classical soft theorem. The large $L$ limit leads to the $\tau$ and $\rho$ coordinates being scaled down to the locally flat spacetime patch with respective coordinates $t$ and $r$ following \ref{t.tran}. However, angular separations between points on the AdS$_4$ boundary and the flat spacetime patch are not necessarily small as would be the case in using isotropic coordinates. We expand on this point and make it precise below. 

To find the perturbed flat spacetime soft theorem from the CFT$_3$ Ward identity, we follow the treatment in Sec.~\ref{sec2} with $\alpha(x')$ in \ref{gpa.ads} defined in terms of $\epsilon^{\text{L}}(\hat{x}'')$. Noting that the map between correlation functions of primary operators and $S$-matrix elements in the $L \to \infty$ limit is not affected by our analysis, we find that the CFT$_3$ Ward identity provides the following $1/\gamma^2$ corrected soft photon theorem  due to $1/L^2$ corrections to the soft photon mode

\begin{equation}
\lim_{\omega_q \rightarrow 0} \frac{\sqrt{2} \omega_q }{\left(1 + z_q \bar{z}_q\right)}\langle \text{out}\vert \hat{a}_{\vec{q}}^{\text{out; L}\, (+)}(\omega_q \hat{x}) \mathcal{S} \vert \text{in} \rangle =  \left[\ \sum_{k = \text{out}} \epsilon^{\text{L}}(x') Q_{k}  - \sum_{k=\text{in}} \epsilon^{\text{L}}(x') Q_k \right] \langle \text{out}\vert \mathcal{S} \vert \text{in} \rangle \,,
\label{spt.ads}
\end{equation}

where we have made use of the $CPT$ invariance of matrix elements in the in-state and out-state to arrive at the result in \ref{spt.ads}. 

The expression in \ref{spt.ads} has a gauge parameter that involves an integration over intermediate angles and hence is not the same as the result derived from classical soft theorems in \ref{spt.cst}.  We also note that by evaluating \ref{atoj.L} as a contour integral with higher order poles located at $z_q$ and $\bar{z}'$, we get a result with a delta function that relates $z_q$ with $z'$, which would violate our assumption of a fixed current on the AdS$_4$ boundary. Thus we need to proceed differently to extract a gauge parameter expression with no dependence on intermediate angular coordinates just as in the classical soft photon theorem result.

One way to identify a gauge parameter expression that only depends on $\{z_q\,,\bar{z}_q\}$ and $\{z'\,,\bar{z}'\}$ is to consider the distance $\vert z_q - z' \vert \approx \tilde{\epsilon}$ as the smallest regulated length scale, with $\{z_w\,,\bar{z}_w\}$ separated from either $\{z_q\,,\bar{z}_q\}$ or $\{z'\,,\bar{z}'\}$ with the expansion

\begin{equation}
    z_w = z_q + \delta e^{i \theta} \,, \quad z_w = z' + \delta e^{i \theta}\,,
    \label{zw.exp}
\end{equation}
We will not particularly distinguish the modulus $\delta$ and phase $\theta$ in the two expansions, since $\{z_q\,,\bar{z}_q\}$ and $\{z'\,,\bar{z}'\}$ are considered close to one another.

Before proceeding, we make a few comments on this approximation. On the one hand, we can consider it as a means of regulating the delta function answer that would result from integrating over $\{z_{w}\,,\bar{z}_{w}\}$ in \ref{gpa.adsc}. In another way the above consideration brings us on similar footing as achieved by the choice of isotropic coordinates used in the derivation of the classical soft theorem. By considering an expansion with $\vert z_q - z' \vert$ taken to be the smallest distance, we would then expect to find a leading contribution to the gauge parameter that agrees with the classical soft photon result and a remainder considered as corrections.

In considering \ref{zw.exp} along with $\vert z_q - z'\vert$ as the smallest distance, it follows that the integrand of  the positive helicity gauge parameter in \ref{gpa.adsc} has the leading contribution

\begin{align}
\frac{\left(1+ z' \bar{z}'\right)^2\left(1+ z_w \bar{z}_w\right)^2}{\left(\bar{z}' - \bar{z}_{w}\right)^2 \left(z_q - z_w\right)^3} = \frac{(1+z_q\bar{z}_q)^2 (1+z'\bar{z}')^2}{(\bar{z}_q - \bar{z}')^2 (z_q - z')^3} \left[1 + \mathcal{O}(\delta) \right]\,.
\label{gp.int}
\end{align}

We can formally integrate \ref{gp.int} over $\{z_w\,,\bar{z}_w\}$. Denoting the integration over the $\mathcal{O}(\delta)$ contributions as ``corrections", we find the following result on substituting \ref{gp.int} in the gauge parameter for the positive helicity case in \ref{gpa.adsc}

\begin{align}
    \epsilon^{\text{L}}(\hat{x}') &=\frac{1}{ 2 \gamma^2} \frac{(1+z_q\bar{z}_q)^2 (1+z'\bar{z}')^2}{(\bar{z}_q - \bar{z}')^2 (z_q - z')^3}
     + \text{corrections}\,.
     \label{adssf.fin}
 \end{align}

Hence the $1/L^2$ corrected soft photon mode in \ref{spt.ads} takes the form

\begin{align}
& \lim_{\omega_q \rightarrow 0} \frac{\sqrt{2} \omega_q }{\left(1 + z_q \bar{z}_q\right)}\langle \text{out}\vert \hat{a}_{\vec{q}}^{\text{out; L}\, (+)}(\omega_q \hat{x}) \mathcal{S} \vert \text{in} \rangle \notag\\
 & \qquad =  \frac{1}{2 \gamma ^2}\left[\ \sum_{k = \text{out}} \frac{\left(1 + z' \bar{z}'\right)^2 \left(1 + z_q \bar{z}_q\right)^2}{(\bar{z}_q-\bar{z}')^2 (z_q-z')^3}Q_{k}  - \sum_{k=\text{in}} \frac{\left(1 + z' \bar{z}'\right)^2 \left(1 + z_q \bar{z}_q\right)^2}{(\bar{z}_q -\bar{z}')^2 (z_q - z')^3}Q_k \right] \langle \text{out}\vert \mathcal{S} \vert \text{in} \rangle\notag\\
 &\qquad \qquad \qquad \qquad \qquad \qquad  + \text{corrections}
\label{spt.ads2}
\end{align}

We find that the leading contribution of \ref{spt.ads2} agrees with \ref{spt.cst} on choosing $n_L = \frac{1}{8}$. We recall that while classical soft theorems recover the $1/L^2$ corrected soft photon factor, there remained an overall factor of $n_L$ in the normalization of perturbed soft photon mode. The derivation from AdS/CFT provides a resolution of this ambiguity.

The nature of the corrections in \ref{spt.ads2} in the context of classical soft theorems remain to be better understood. It is clear that the integration over intermediate angles can also be interpreted as a sum over certain particles parametrized by the angular coordinates $\{z_w \,,\bar{z}_w\}$. In this way, while these contributions are present in the AdS/CFT derivation of the $1/L^2$ corrected soft photon mode, they might correspond to excitations in the context of classical soft theorems.

\section{Discussion}

In this paper, we have refined the implications of scattering on non-asymptotically flat spacetimes on known soft theorems. To be precise, we have provided a definition of universal structures in `AdS soft theorems' for field theories on asymptotically AdS spacetimes, with a small cosmological constant. The main result of our paper is the derivation of $1/L^2$ corrections to the flat spacetime soft photon theorem on an AdS$_4$ spacetime from a large $N$ CFT$_3$ Ward identity. This derivation for modes in a locally flat patch of the spacetime made use of bulk gauge fields reconstructed from a $U(1)$ boundary current via the HKLL procedure. We further noted that this result from a CFT$_3$ Ward identity, in a certain limit, recovers our previous result for the corrected soft photon theorem derived from the classical soft photon theorem. Our results hence provide evidence for universal `subleading in AdS radius' corrections to soft theorems satisfied by a $S$-matrix on asymptotically flat spacetimes within a larger AdS spacetime. 

 One aspect of the corrected soft photon mode in \ref{atoj.L} which distinguishes it from the flat spacetime mode in \ref{atoj.flat} is the dependence on intermediate angles. We believe this feature holds to higher orders of AdS radius contributions as well. Through our analysis, we have noted that the HKLL kernels to order $n$ generically appear to have terms with an order $2n$ polynomial of the angular momentum mode $l$. Such terms can be expressed in terms of derivatives on the spherical harmonics with the consequence of additional Green's functions integrated over intermediate angles. Hence the inclusion of intermediate angles at $1/L^2$ appears to be a property that holds to higher orders in the expansion. 
 
 The $1/L^2$ corrected soft theorem derived from a $U(1)$ Ward identity can be considered a complete result for scattering on asymptotically AdS$_4$ spacetimes and more specifically for the $S$-matrix defined on the flat spacetime patch. This raises questions on their relevance in infrared properties of scattering processes. For instance, the Weinberg soft photon factor is the leading infrared divergence coming from real soft photons which cancel out with the infrared divergences coming from photon loop contributions to provide an IR finite scattering processes on asymptotically flat spacetimes. The situation on AdS spacetimes is most likely different, as the AdS radius $L$ is known to be a natural infrared regulator \cite{Callan:1989em,Fitzpatrick:2011jn} providing an exponential decay for massless particles. It is thus tempting to conjecture that the resummation of $1/\omega^n L^n$ corrections to the soft factor (for $n > 2$) leads to emitted massless particles being infrared finite. This remains a topic to explore in the future.
 
 We also noted in Section \ref{match} that the agreement of this result with the classical soft photon theorem results from expanding about a leading saddle independent of intermediate angles. One way to interpret the integration over intermediate angles $\{z_w\,,\bar{z}_w\}$ is that they correspond to additional particles whose momenta are parametrized in terms of these coordinates. We can thus conclude that the classical soft theorem is recovered in a limit that ignores the contributions from these additional particles.  
While our analysis derived the $1/L^2$ corrected soft factor resulting from inserting a soft photon to a $S$-matrix in the flat spacetime patch, it will be important to consider the factorization in $1/L^2$ corrected scattering amplitudes such as those recently derived in \cite{Komatsu:2020sag, Li:2021snj}.  Given the universality of the leading soft factor, including $1/L^2$ corrections, this should be derivable for these amplitudes as well.

\section{Acknowledgments}
We would like to thank Yu-tin Huang, Heng-Yu Chen and Hikaru Kawai for discussions and valuable feedback on our results. KF is supported by the Ministry of Science and Technology (MOST), Taiwan through the grant MOST 111-2811-M-003-005 and would like to thank Harish-Chandra Research Institute, National Taiwan University and National Taiwan Normal University for their hospitality during the completion of this work. The work of AM was supported by the National Science and Technology Council, the Ministry of Education (Higher Education Sprout Project NTU-111L104022), and the National Center for Theoretical Sciences of Taiwan. AM would like to thank IISER Bhopal for their hospitality during the initial stage of this work.

\appendix

\section{Solution of Maxwell's equations in AdS$_4$}\label{appa}

We will be interested in solutions of Maxwell's equations in the absence of sources

\begin{equation}
\nabla^{\mu} \mathcal{F}_{\mu \nu} = 0 \,,
\label{max.ads}
\end{equation}

where $\mathcal{F}_{\mu \nu} = \nabla_{\mu} \mathcal{A}_{\nu} - \nabla_{\nu} \mathcal{A}_{\mu}$ with $\mathcal{A}_{\mu}$ the bulk gauge field,  $\nabla^{\mu}$ the covariant derivative with respect to the background. We will follow the treatment by Wald and Ishibashi~\cite{Ishibashi:2004wx} in deriving the classical solutions. The general metric 
\begin{equation}
ds^2 = g_{\mu\nu}dx^{\mu} dx^{\nu} = h_{ab}dy^a dy^b + \tilde{g}_{ij}dz^i dz^j \label{ads4.gen}
\end{equation}
can be expressed in the global form of \ref{ads.met2} by choosing
\begin{align}
h_{ab} &= \frac{L^2}{\cos^2\rho} \eta_{ab} \notag\\
\tilde{g}_{ij}dz^i dz^j &=   \frac{L^2}{\cos^2 \rho} \sin^2 \rho \frac{4}{(1+z \bar{z})^2}dz d\bar{z} = 2 L^2 \tan^2 \rho \gamma_{z \bar{z}} dz d\bar{z} \label{met.spher}
\end{align}

We carry out a vector harmonic decomposition of the Maxwell field into the following independent components
\begin{align}
\mathcal{A}^V_{i}dx^{i} &= \sum_{l,m} \Psi^{lm}(\tau\,,\rho) \epsilon_{ij}\partial^j Y_{lm}dz^i \,, \label{m.v}\\
\mathcal{A}^S_{\mu}dx^{\mu} &= \sum_{l,m} \left[A^{lm}_a (\tau\,,\rho) Y_{lm} dy^a + A^{lm}(\tau\,,\rho) \partial_{i}Y_{lm}dz^i\right] \,,  \label{m.s}
\end{align}

The superscripts $V$ and $S$ respectively refer to vector and scalar type components. The scaling dimensions of the two components differ : $\Delta = 2$ in the vector case and $\Delta = 1$ in the scalar case.

From Maxwell's equations \ref{max.ads} we find that the vector type component in \ref{m.v} manifestly satisfies 

\begin{equation}
\Box \Psi^{lm} - \frac{l(l+1)}{\sin^2\rho}\Psi^{lm} = 0 \label{maxr.sol}
\end{equation}

The scalar type component can also be shown to satisfy a similar equation by defining the field `$\phi^{lm}$' constructed from $A^{lm}_a$ and $A^{lm}$ in the following way

\begin{equation}
\partial_a A^{lm} - A^{lm}_a = \epsilon_{ab} \partial^b \phi^{lm}(\tau\,,\rho) \,, \label{maxe.lc}
\end{equation}

On substituting the scalar type expression \ref{m.s} in \ref{max.ads}, we find  

\begin{equation}
\Box \phi^{lm} - \frac{l(l+1)}{\sin^2\rho}\phi^{lm} = 0 \label{maxs.sol}
\end{equation}

The solutions we need are those that satisfy the HKLL asymptotic matching condition

\begin{equation}
j_{\mu} = \lim_{\cos \rho \to \frac{\pi}{2}} \cos^{-1} \rho \mathcal{A}_{\mu}
\label{hkllmap}
\end{equation}

Thus for purely radiative solutions derived in the absence of any $j_{a}$ current component (no Coulombic fields), the contribution from $A^{lm}_a$ in \ref{maxe.lc} drops out of the scalar type solution. The resulting equation \ref{maxs.sol} simplifies to

\begin{equation}
\Box A^{lm} - \frac{l(l+1)}{\sin^2 \rho} A^{lm} = 0 
\label{maxa.sol}
\end{equation}

which is the same as the vector type equation \ref{maxr.sol}. We will henceforth denote $A^{lm}$ in \ref{maxa.sol} and $\Psi^{lm}$ in \ref{maxr.sol} commonly by $\Phi^{lm}$, with the solutions distinguished by different values of $\Delta$. The solution of the radial equation \ref{maxr.sol} and \ref{maxa.sol} is  

\begin{align} 
\Phi^{lm}(\tau\,,\rho) &= e^{\pm i \omega_{\kappa} \tau}\Phi^{lm}(\rho) \label{scal.sol}\\
\text{with} \quad \Phi^{lm}(\rho) & \sim  \sin^{l+1} \rho \cos^{\Delta-1} \rho \,_2F_1\left(- \kappa \,, \kappa + \Delta + l \,, \Delta - \frac{1}{2}\Big\vert \cos^2 \rho \right) \label{scal.tot}\\
& \qquad \text{where} \quad \kappa = \frac{\omega_{\kappa} - \Delta - l}{2}
\end{align} 

The $\sim$ in \ref{scal.tot} indicates an as yet unspecified overall normalization.

A feature of the $\{z\,,\bar{z}\}$ coordinates is that the derivative basis simplifies considerably 
\begin{align}
\partial_{i}Y_{lm} & = \partial_{z} Y_{lm} \qquad (\text{for}\; i = z) \;; \qquad  \partial_{i}Y_{lm} = \partial_{\bar{z}} Y_{lm} \qquad (\text{for}\;  i = \bar{z}) \label{even.sph}\\
\epsilon_{ij}\partial^j Y_{lm}& = \partial_{z} Y_{lm} \qquad (\text{for}\;  i = z)\;; \qquad \epsilon_{ij}\partial^j Y_{lm} = -\partial_{\bar{z}} Y_{lm} \qquad (\text{for}\;  i = \bar{z}) \label{odd.sph}
\end{align}

Hence the classical solutions that enter our analysis are simply
\begin{align}
\mathcal{A}^V_{z}(\tau\,,\rho\,,\Omega) &\sim \sum_{l,m} \Phi^{lm}(\tau\,,\rho)\Big\vert_{\Delta=2}\partial_z Y_{lm}(\Omega) \;; \qquad \mathcal{A}^V_{\bar{z}}(\tau\,,\rho\,,\Omega) \sim  \sum_{l,m} - \Phi^{lm}(\tau\,,\rho)\Big\vert_{\Delta=2}\partial_{\bar{z}} Y_{lm}(\Omega) \notag\\
\mathcal{A}^S_{z}(\tau\,,\rho\,,\Omega) &\sim \sum_{l,m} \Phi^{lm}(\tau\,,\rho)\Big\vert_{\Delta=1}\partial_z Y_{lm}(\Omega) \;; \qquad \mathcal{A}^S_{\bar{z}}(\tau\,,\rho\,,\Omega) \sim  \sum_{l,m} \Phi^{lm}(\tau\,,\rho)\Big\vert_{\Delta=1}\partial_{\bar{z}} Y_{lm}(\Omega) \notag
\end{align}

These solutions, along with the general time dependence $e^{i \omega_{\kappa} (\tau - \tau')}$, define the function $\Xi_{\kappa l}(\rho,\tau,\tau')$ in the vector and scalar type kernels of \ref{kv.1} - \ref{ks.2}.

\section{$1/L^2$ corrections of the gauge field HKLL kernels} \label{appb}

We will now describe the derivation of the $1/L^2$ corrected expressions for $\Xi_{\kappa l}(\rho,\tau,\tau')$ in \ref{xi.d2} and \ref{xi.d1}, from the general expression given in \ref{xi.def}, which we repeat here for convenience

\begin{equation}
\Xi_{kl}(\rho\,,\tau\,,\tau') = e^{i \omega_{\kappa} (\tau - \tau')} \sin^{l+1} \rho \cos^{\Delta-1} \rho \,_2F_1\left(- \kappa \,, \kappa + \Delta + l \,, \Delta - \frac{1}{2}\Big\vert \cos^2 \rho \right)\,,
\label{xi.defa}
\end{equation}

We will describe three intermediate steps leading to a form of $\Xi_{\kappa l}$ that we consider. The first is the transformation of $\cos^2 \rho$ to  $\sin^2 \rho$ in the hypergeometric function argument by a linear transformation (cf. 2.4 of \cite{MO:1966})  \footnote{We consider this transformation since $1/L^2$ corrections of flat spacetime still involve Bessel functions (and their derivatives) with the argument $r \omega$. It is simpler to recover these Bessel functions from a $\sin^2 \rho$ argument in the hypergeometric function.}. On transforming the hypergeometric function, we find a coefficient with products of Gamma functions, some of which involve a negative argument. These can be transformed to a positive argument, and we specifically consider

\begin{equation}
\frac{\Gamma(-l - \frac{1}{2})}{\Gamma(-\kappa - l - \frac{1}{2})} = (-1)^{-\kappa} \frac{\Gamma(\kappa + l + \frac{3}{2})}{\Gamma(l + \frac{3}{2})} \,,
\label{gam.eul}
\end{equation}

which is derived from the Euler reflection identity $\Gamma(x) \Gamma(1-x) = \frac{\pi}{\sin \pi x} $ (for non-integer $x$). Lastly, we replace $\omega_{\kappa} = \omega L$ and $\kappa = \frac{1}{2}( - \omega L + \Delta + l)$, with $\kappa$ considered large.

The resulting expression for $\Xi_{\kappa l}$ is

\begin{align}
\Xi_{\kappa l}(\rho\,,\tau\,,\tau') &= (\pm i)^{-\Delta - l} e^{ \pm i \omega L \frac{\pi}{2}} e^{i \omega L (\tau - \tau')} \times A \times B\times C \notag\\
\text{where}\quad & A = \tan^{l+1} \rho \cos^{\Delta+l} \rho \notag\\
& B = \frac{\Gamma\left(\Delta - \frac{1}{2}\right)\Gamma\left(\frac{\omega L +l - \Delta + 3}{2}\right)}{\Gamma(l + \frac{3}{2}) \Gamma\left(\frac{\omega L - l + \Delta -1}{2}\right)} \notag\\
& C = \,_2F_1\left(\frac{\Delta + l - \omega L}{2} \,, \frac{\Delta + l + \omega L}{2} \,, l + \frac{3}{2}\Big\vert \sin^2 \rho \right)
\label{xi.part}
\end{align}

The $(\pm i)^{-\Delta - l} e^{ \pm i \omega L \frac{\pi}{2}}$ comes from the $(-1)^{-\kappa}$ in \ref{gam.eul}. The $+$ ($-$) sign will represent positive frequency outgoing (incoming) states in the kernels. \footnote{The converse convention holds for negative frequency states and follows from complex conjugation.}

The $1/L^2$ corrections to the flat spacetime result from the HKLL kernels will result from expanding the above terms after substituting \ref{t.tran}.  As $\tau$ involves a trivial rescaling, we find 

$$e^{i \omega L (\tau - \tau')} = e^{i \omega t} e^{-i \omega L \tau'} \,,$$

just as in the flat spacetime limit, which holds to all orders in $1/L^2$. The non-trivial expansions in $1/L^2$ come from the terms $A$, $B$ and $C$ noted above on replacing $\rho = \arctan(\frac{r \omega}{\omega L})$. Performing a Taylor expansion on $A$ gives the following result

\begin{align}
A = \tan^{l+1} \rho \cos^{\Delta+l} \rho &=  \left(\frac{r \omega}{\omega L}\right)^{l+1} \left[1 - \frac{(l+ \Delta)(r \omega)^2}{2 \omega^2 L^2} + \mathcal{O}\left(\frac{1}{\omega^3 L^3}\right)\right] \label{a.1}
\end{align}

In the case of $B$, and specifically for the factor $\frac{\Gamma\left(\frac{\omega L +l - \Delta + 3}{2}\right)}{\Gamma\left(\frac{\omega L - l + \Delta -1}{2}\right)}$ appearing within it, we can make use of the following identity for Gamma functions with a large argument $x$
(cf. 5.11.13 of \cite{NIST})

\begin{align}
\frac{\Gamma(x+a)}{\Gamma(x+b)} &= x^{a-b}\left( 1 + \frac{1}{2}\frac{(a-b)(a+b-1)}{x} \right.\notag\\
&\left. \qquad \qquad + \frac{1}{12}\frac{(a-b)(a-b-1)(3(a+b-1)^2 - (a-b+1))}{2 x^2} + \mathcal{O}(x^{-3})\right)
\label{gam.frac}
\end{align}

As $\omega L$ is large, we define
$x=\frac{\omega L}{2}$ , $a= \frac{l+3-\Delta}{2}$ and $b= \frac{\Delta-l-1}{2}$ to find

\begin{equation}
\frac{\Gamma(\frac{\omega L +l - \Delta + 3}{2})}{\Gamma(\frac{\omega L - l + \Delta -1}{2})} = \left(\frac{\omega L}{2}\right)^{l+2-\Delta}\left(1 - \frac{(l+1 - \Delta)(l+2 - \Delta)(l+3 - \Delta)}{6\omega^2 L^2} + \mathcal{O}\left(\frac{1}{\omega^3 L^3}\right)\right)
\label{b.part}
\end{equation}

Thus our expansion for the $B$ term is
\begin{align}
B =\frac{\Gamma(\Delta-\frac{1}{2})}{\Gamma(l+\frac{3}{2})} \left(\frac{\omega L}{2}\right)^{l+2-\Delta}\left(1 - \frac{(l+1 - \Delta)(l+2 - \Delta)(l+3 - \Delta)}{6\omega^2 L^2} + \mathcal{O}\left(\frac{1}{\omega^3 L^3}\right)\right)
\label{b.1}
\end{align}

For the $C$ term, we make use of the following expansion of the hypergeometric function in terms of Bessel functions~\cite{Thorsley:2001}

\begin{align}
 \,_2F_1\left(\lambda\,, \mu \,; \nu+1 \Bigg\vert -\frac{y^2}{4 \lambda \mu}\right)  = \Gamma(\nu + 1)\left(\frac{y}{2}\right)^{-\nu} &\left[ J_{\nu}(y) + \frac{y^2}{8}J_{\nu+2}(y) \left(\frac{1}{\lambda} + \frac{1}{\mu}\right) \phantom{\Bigg\vert}\right.\notag\\
&\left. \quad  + \left[\frac{y^4}{128}J_{\nu+4}(y) - \frac{y^3}{24}J_{\nu+3}(y)\right] \left(\frac{1}{\lambda^2} + \frac{1}{\mu^2}\right) \right. \notag\\
&\left. \qquad + \left[\frac{y^4}{64}J_{\nu+4}(y) - \frac{y^3}{8}J_{\nu+3}(y) + \frac{y^2}{8}J_{\nu+2}(y) \right] \frac{1}{\lambda \mu} \right.\notag\\
&\left. \quad \phantom{\Bigg\vert} + \mathcal{O}(\lambda^{-3}\,,\mu^{-3}\,,\lambda^{-1}\mu^{-2}\,, \cdots) \right]
\label{2f1.exp} 
\end{align}

The leading contribution in \ref{2f1.exp} is the relationship between the hypergeometric and Bessel functions derived by Watson~\cite{watson}. The expansion in \ref{2f1.exp} was determined through Watson's approach carried out to subleading order~\cite{Thorsley:2001} and will be needed to determine the $1/L^2$ corrections of the HKLL kernels. 

On comparing the expression for $C$ in \ref{xi.part} with \ref{2f1.exp}, we find 

\begin{equation} 
\lambda = \frac{\Delta + l - \omega L}{2}\,, \qquad \mu = \frac{\Delta + l + \omega L}{2}
\label{lm}
\end{equation}

We can likewise determine $y^2$ in \ref{2f1.exp} from $\sin^2 \rho$ in $C$. Since 

\begin{equation}
-4 \lambda \mu = \omega^2 L^2 - \left(\Delta + l \right)^2\,,
\label{lm.prod}
\end{equation}

we can appropriately replace $\omega^2 L^2 $ with $-4 \lambda \mu$ in the expansion for $\sin^2 \rho$ to find
\begin{align}
\sin^2 \rho &=  \left(\frac{r \omega}{\omega L}\right)^2 \left(1 - \frac{r^2 \omega^2}{\omega^2 L^2} + \mathcal{O}\left(\frac{1}{\omega^3 L^3}\right)\right) \notag\\
&= \frac{r^2 \omega^2}{-4 \lambda \mu} \left(1 - \frac{\left(\Delta + l \right)^2 + r^2 \omega^2}{\omega^2 L^2} \right) + \mathcal{O}\left(\frac{1}{\omega^3 L^3}\right) \notag\\
& := \frac{y^2}{-4 \lambda \mu} + \mathcal{O}\left(\frac{1}{\omega^3 L^3}\right)\,,
\label{y.rel}
\end{align}

where in the last line of \ref{y.rel} we defined

\begin{equation}
y = r\omega \left(1 - \frac{\left(\Delta + l\right)^2 + r^2 \omega^2}{2 \omega^2 L^2} \right)
\label{y.def}
\end{equation}

We can hence derive the right hand side of \ref{2f1.exp} from the given expression of $C$ in \ref{xi.part}. Each Bessel function appearing in the expression can be written in terms of the flat spacetime argument $r \omega$ by making use of 

\begin{equation}
J_{\nu}(x + \delta x) = J_{\nu}(x) - \delta x J_{\nu + 1}(y) + \frac{\nu}{x} J_{\nu}(x)\,,
\label{bess.rec}
\end{equation}

which can be derived from recursion relations for the Bessel functions

\begin{align}
J_{\nu + 1}(x) &= -J'_{\nu}(x) + \frac{\nu}{x} J_{\nu}(x)\,, \label{rec}\\
J'_{\nu + 1}(x) &= J_{\nu}(x) - \frac{\nu+ 1}{x} J_{\nu+1}(x)\,, \label{drec}
\end{align}

In particular, on using \ref{y.def} we find that \ref{bess.rec} gives

\begin{equation}
J_{\nu}(y) = J_{\nu}(r \omega) + \frac{\left(\Delta + l\right)^2 + r^2 \omega^2}{2 \omega^2 L^2}\left(r \omega J_{\nu + 1}(r \omega) - \nu J_{\nu}(r \omega)\right)
\label{bess.exp}
\end{equation}

On replacing \ref{lm} and \ref{y.rel} in \ref{2f1.exp}, and  expressing all the Bessel function arguments in terms of $r \omega$, we find the following expression for $C$

\begin{align}
C &= \Gamma\left(l+\frac{3}{2}\right) \left(\frac{2}{r\omega}\right)^{l+\frac{1}{2}} \left[J_{l+\frac{1}{2}}(r \omega) + \frac{r \omega \left(\left(\Delta + l\right)^2 + r^2 \omega^2\right)}{2 \omega^2 L^2} J_{l + \frac{3}{2}}(r \omega)  \right. \notag\\
&\left. \phantom{\Bigg\vert} \qquad \qquad \qquad \qquad \qquad - \frac{r^2 \omega^2 \left(l + \Delta + 1\right)}{2 \omega^2 L^2} J_{l+\frac{5}{2}}(r \omega)  + \frac{r^3 \omega^3}{6 \omega^2 L^2} J_{l+ \frac{7}{2}}(r \omega)\right]
\label{c.1}
\end{align}

We can now multiply \ref{a.1}, \ref{b.1} and \ref{c.1} to get the expression for $\Xi_{\kappa l}(r,t ; \tau')$ from \ref{xi.part}. Further simplifications can be performed -- the first involves the use of recursion relations for Bessel functions in \ref{rec} and \ref{drec}, which enable finding an expression involving only $J_{l+\frac{1}{2}}(r \omega)$ and its first derivative. Each $J_{l+\frac{1}{2}}(r \omega)$ can then be written in terms of Spherical Bessel functions $j_{l}(r\omega)$

\begin{equation}
j_l(r \omega) = \sqrt{\frac{\pi}{2 r \omega}} J_{l+\frac{1}{2}}(r \omega)\,.
\label{sb.def}
\end{equation}

The other simplification that occurs is for a common expression in the two cases $\Delta = 1$ and $\Delta = 2$. We specifically have

\begin{equation}
\Gamma\left(\Delta - \frac{1}{2}\right) 2^{\Delta - 1}\Bigg\vert_{\Delta=2} =  \quad \sqrt{\pi} \quad  = \Gamma\left(\Delta - \frac{1}{2}\right) 2^{\Delta - 1}\Bigg\vert_{\Delta=1}
\end{equation}

Following the use of Bessel function recursion relations and the substitutions mentioned above, we then find

\begin{align}
\Xi_{\kappa l}(\rho\,,\tau\,,\tau')\Big \vert_{\Delta = 2} &= - (\pm i)^{- l} e^{i \omega t} e^{-i \omega L \left( \tau' \mp \frac{\pi}{2}\right)}\left[A \times B \times C\right]_{\Delta = 2}\,, \label{xi.db2}\\
\Xi_{\kappa l}(\rho\,,\tau\,,\tau')\Big \vert_{\Delta = 1} &= - (\pm i)(\pm i)^{- l} e^{i \omega t} e^{-i \omega L \left( \tau' \mp \frac{\pi}{2}\right)}\left[A \times B \times C\right]_{\Delta = 1}\,,  \label{xi.db1}
\end{align}
with 
\begin{align}
\left[A \times B \times C\right]_{\Delta = 2}  &= \frac{r}{L}  \left[ j_{l}(r \omega) + \frac{1}{2 \omega^2 L^2}\left(\left(\frac{l(l+1)}{2} - \frac{(r \omega)^2}{3}\right)j_{l}(r \omega)  \right. \right. \notag\\
&\left. \left. \phantom{\frac{1}{2 \omega^2 L^2}}- \frac{2 r\omega}{3} \sqrt{\frac{\pi}{2 r \omega}} \left(\frac{l (l+1)}{2} +  (r \omega)^2 \right) J'_{l+\frac{1}{2}}(r \omega)\right) \right]  + \mathcal{O}\left(\frac{1}{\omega^3 L^3}\right) \label{abc.d2}\\
\left[A \times B \times C\right]_{\Delta = 1}  &=  \frac{r}{L} (\omega L) \left[ j_{l}(r \omega) - \frac{1}{2 \omega^2 L^2}\left(\left(\frac{l(l+1)}{2} + \frac{(r \omega)^2}{3}\right) j_{l}(r \omega)  \right. \right. \notag\\
&\left. \left. \phantom{\frac{1}{2 \omega^2 L^2}}\qquad  - \frac{2 r\omega}{3} \sqrt{\frac{\pi}{2 r \omega}} \left(\frac{l (l+1)}{2} +  (r \omega)^2 \right) J'_{l+\frac{1}{2}}(r \omega)\right) \right]  + \mathcal{O}\left(\frac{1}{\omega^3 L^3}\right) \label{abc.d1}
\end{align}
The expressions in \ref{xi.db2} and \ref{xi.db1} are those in \ref{xi.d2} and \ref{xi.d1} respectively.

\section{Derivation of $\hat{a}_{\vec{q}}^{\text{out; L}\, (+)}$ and $\hat{a}_{\vec{q}}^{\text{out}(+)}$} \label{appc}
The outgoing positive helicity photon modes result from substituting the outgoing bulk expression for $\hat{{\cal A}}^{\text{out}}_{z}(y)$ from \ref{az.full} in \ref{an.fl}.  We define the flat spacetime mode $\hat{a}_{\vec{q}}^{\text{out}(+)}$ as that corresponding to the bulk field $\hat{{\cal A}}^{\text{out; f}}_{z}(y)$ and the $1/L^2$ corrected mode $\hat{a}_{\vec{q}}^{\text{out; L}\, (+)}$ as that resulting from the bulk field $\hat{{\cal A}}^{\text{out; L}}_{z}(y)$ in the following way  
\begin{align}
\hat{a}_{\vec{q}}^{\text{out}\, (+)} &= \lim_{t\rightarrow \infty} i \int d^3 \vec{y} \,  (\varepsilon^{(+) \mu})^*  e^{-i q\cdot y} \overleftrightarrow{\partial_0} \hat{{\cal A}}^{\text{out; f}}_{z}(y)\,,
    \label{aof}\\
   \hat{a}_{\vec{q}}^{\text{out; L}\, (+)} &= \lim_{t\rightarrow \infty} i \int d^3 \vec{y} \,  (\varepsilon^{(+) \mu})^*  e^{-i q\cdot y} \overleftrightarrow{\partial_0} \hat{{\cal A}}^{\text{out; L}}_{z}(y)\,,
    \label{aol}
\end{align}
In both cases, we use the expressions for the polarization and plane waves given in \ref{pl.exp1}. On substituting $\hat{{\cal A}}^{\text{out; f}}_{z}(y)$ from \ref{az.flat} and $\hat{{\cal A}}^{\text{out; L}}_{z}(y)$ from \ref{az.ads}, we then find that the expressions in \ref{aof} and \ref{aol} take the form

\begin{align}
\hat{a}_{\vec{q}}^{\text{out}\, (+)} &=  \lim_{t\rightarrow \infty} i \int r^2 dr \, \int d\Omega \, \int\limits_{0}^{\pi} d\tau' \int d\Omega' \int d\omega \, \frac{1 + z \bar{z}}{\sqrt{2}} \sum_{l,m,l',m'} j_{l'}(r \omega_{\vec{q}}) j_{l}(r \omega) \notag\\
& \; \left[ (i (\omega - \omega_q)) \frac{Y^*_{lm}\left(\Omega'\right)}{-l(l+1)} Y_{l'm'}(\Omega) Y^*_{l'm'}(\Omega_{q}) \partial_{z}Y_{lm}\left(\Omega \right) (-i)^{l'} (i)^{- l} e^{i (\omega + \omega_{\vec{q}}) t} e^{-i \omega L \left( \tau' - \frac{\pi}{2}\right)}\, D^{\bar{z}'} j^+_{\bar{z}'} \right.\notag\\
&\left. \quad  + (-i (\omega + \omega_{\vec{q}}))\frac{Y_{lm}\left(\Omega'\right)}{-l(l+1)} Y_{l'm'}(\Omega) Y^*_{l'm'}(\Omega_{q}) \partial_{z}Y^*_{lm}\left(\Omega \right) (- i)^{- l + l'} e^{-i (\omega - \omega_{\vec{q}}) t} e^{i \omega L \left( \tau' - \frac{\pi}{2}\right)} D^{\bar{z}'} j^-_{\bar{z}'} \right] \label{aof.pexp}\\
\hat{a}_{\vec{q}}^{\text{out; L}\, (+)} &=  \lim_{t\rightarrow \infty} i \int r^2 dr \, \int d\Omega \, \int\limits_{0}^{\pi} d\tau' \int d\Omega' \int d\omega \frac{1 + z \bar{z}}{\sqrt{2}} \sum_{l,m,l',m'} j_{l'}(r \omega_{\vec{q}}) j_{l}(r \omega) \left(\frac{l(l+1)}{4 \omega^2 L^2}\right)\notag\\
& \; \left[ (i (\omega - \omega_q)) \frac{Y^*_{lm}\left(\Omega'\right)}{-l(l+1)} Y_{l'm'}(\Omega) Y^*_{l'm'}(\Omega_{q}) \partial_{\bar{z}}Y_{lm}\left(\Omega \right) (-i)^{l'} (i)^{- l} e^{i (\omega + \omega_{\vec{q}}) t} e^{-i \omega L \left( \tau' - \frac{\pi}{2}\right)}\, D^{\bar{z}'} j^+_{\bar{z}'} \right.\notag\\
&\left. \quad  + (-i (\omega + \omega_{\vec{q}}))\frac{Y_{lm}\left(\Omega'\right)}{-l(l+1)} Y_{l'm'}(\Omega) Y^*_{l'm'}(\Omega_{q}) \partial_{\bar{z}}Y^*_{lm}\left(\Omega \right) (- i)^{- l + l'} e^{-i (\omega - \omega_{\vec{q}}) t} e^{i \omega L \left( \tau' - \frac{\pi}{2}\right)} D^{\bar{z}'} j^-_{\bar{z}'} \right] \label{aol.pexp}
\end{align}

The above expressions contain derivatives of spherical harmonics and in this regard, it is useful to introduce the Green's function $G(z\,, \bar{z}\,; w \,, \bar{w})$ on the $2$-sphere

\begin{equation}
    G(z\,, \bar{z}\,; w \,, \bar{w}) = \frac{1}{4\pi} \ln \left((z - w) (\bar{z} - \bar{w})\right) - \frac{1}{4\pi} \ln \left( 1 + z\bar{z}\right)- \frac{1}{4\pi} \ln \left( 1 + w\bar{w}\right)\,,
    \label{G.sph}
\end{equation}

From the identity
$$ \partial_z\frac{1}{\bar{z} - \bar{w}} = \frac{1}{2} \delta^{(2)}(z - w) = \partial_{\bar{z}}\frac{1}{z - w} $$

we find that $G(z\,, \bar{z}\,; w \,, \bar{w})$ satisfies the following relations

\begin{align}
     \partial_{z} \partial_{\bar{z}} G(z\,, \bar{z}\,; w \,, \bar{w}) =  \frac{1}{2}\delta^{(2)}(z - w) - \frac{\gamma_{z \bar{z}}}{8 \pi} \,, &\qquad  \partial_{w} \partial_{\bar{w}} G(z\,, \bar{z}\,; w \,, \bar{w}) = \frac{1}{2}\delta^{(2)}(z - w) - \frac{\gamma_{w \bar{w}}}{8 \pi}\label{gfs.id}\\
     \partial_{z} \partial_{\bar{w}} G(z\,, \bar{z}\,; w \,, \bar{w}) &=  - \frac{1}{2}\delta^{(2)}(z - w) = \partial_{w} \partial_{\bar{z}} G(z\,, \bar{z}\,; w \,, \bar{w}) \label{gfs.md}
\end{align}

\ref{gfs.md} in particular implies the useful property

\begin{align}
    \partial_{z_p}\frac{Y_{lm}\left(\Omega_{z_p}\right)}{-l(l+1)} &= -2 \int d \Omega_w \gamma^{w \bar{w}} \partial_{\bar{w}} \partial_{z_p} G(z_p\,, \bar{z}_p\,; w \,, \bar{w}) \partial_{w}\frac{Y_{lm}\left(\Omega_{w}\right)}{-l(l+1)} \notag\\
    &= \int d \Omega_w  \partial_{z_p} G(z_p\,, \bar{z}_p\,; w \,, \bar{w}) Y_{lm}\left(\Omega_{w}\right) \,,
    \label{y.md}
\end{align}

where we made use of $2 \gamma^{w \bar{w}} \partial_{w} \partial_{\bar{w}} Y_{lm}\left(\Omega_{w}\right) = - l (l+1) Y_{lm}\left(\Omega_{w}\right)$ in the second equality. 

We can now use the orthogonality relations satisfied by the spherical harmonics

\begin{align}
\int d\Omega Y_{lm}\left(\Omega\right) Y^*_{l'm'}\left(\Omega\right) &= \delta_{ll'} \delta_{mm'} \label{ylm.int}\\
\sum_{l,m} Y_{lm}\left(\Omega_{q}\right)Y^*_{lm}\left(\Omega\right) &= \delta \left(\Omega_{q}- \Omega\right) \,, \label{ylm.orth}
\end{align}

and the spherical Bessel functions

\begin{align}
\int\limits_{0}^{\infty} r^2 dr j_{l}(r \omega) j_{l}(r \omega_{\vec{q}}) &= \frac{\pi}{2 \omega^2_{\vec{q}}} \delta(\omega - \omega_{\vec{q}}) \,, \label{bess.int}
\end{align}

to simplify the integrals appearing in \ref{aof.pexp} and \ref{aol.pexp}. In the case of \ref{aof.pexp} we find the result

\begin{align}
    \hat{a}_{\vec{q}}^{\text{out}\, (+)} &= \pi \frac{1+z_q \bar{z}_q}{\sqrt{2} \omega_q} \int\limits_{0}^{\pi} d\tau' \int d\Omega' \partial_{z_q} G(z_q\,,z') \mathcal{D}^{\bar{z}'}j^{-}_{\bar{z}'} e^{i \omega_q L \left( \tau' - \frac{\pi}{2}\right)} \notag\\
        &= \frac{1}{4} \frac{1+z_q \bar{z}_q}{\sqrt{2} \omega_q} \int\limits_{0}^{\pi} d\tau' \int d\Omega' \frac{1}{z_q - z'} \mathcal{D}^{\bar{z}'}j^{-}_{\bar{z}'} e^{i \omega_q L \left( \tau' - \frac{\pi}{2}\right)}\,,
        \label{aof.pexp1}
\end{align}

which is the flat spacetime mode expression \ref{atoj.flat} that was derived in \cite{Hijano:2020szl}. 

We find that the $1/L^2$ corrected mode in \ref{aol.pexp} simplifies to

\begin{align}
\hat{a}_{\vec{q}}^{\text{out; L}\, (+)} &=  \frac{\pi}{\omega_q^2 L^2} \frac{1 + z_{q} \bar{z}_{q}}{\sqrt{2} \omega_{\vec{q}}} \int\limits_{0}^{\pi} d\tau' \int d\Omega' \int d\Omega_w \notag\\
& \qquad \qquad  \partial_{z_q} \left[ \left(\gamma^{z' \bar{z}'} \partial_{\bar{z}'}\partial_{w} G(z'\,, \bar{z}'\,; w \,, \bar{w})\right) \left(\gamma^{w \bar{w}} \partial_{\bar{w}}\partial_{z_q} G(w\,, \bar{w}\,; z_q \,, \bar{z}_q)\right)\right] D^{\bar{z}'} j^-_{\bar{z}'} e^{i \omega_q L \left( \tau' - \frac{\pi}{2}\right)} \notag\\
&=  \frac{1}{32 \pi \omega_q^2 L^2} \frac{1 + z_{q} \bar{z}_{q}}{\sqrt{2} \omega_{\vec{q}}} \int\limits_{0}^{\pi} d\tau' \int d\Omega' \int d\Omega_w \left[\frac{\left(1+ z' \bar{z}'\right)^2\left(1+ z_w \bar{z}_w\right)^2}{\left(\bar{z}' - \bar{z}_{w}\right)^2 \left(z_q - z_w\right)^3}\right] \mathcal{D}^{\bar{z}'}j^{-}_{\bar{z}'} e^{i \omega_q L \left( \tau' - \frac{\pi}{2}\right)}
\label{aol.pexp1}
\end{align}

which is the expression in \ref{atoj.L}. One key difference between the flat spacetime mode in \ref{aof.pexp1} and the $1/L^2$ corrected mode in \ref{aol.pexp1} is the presence of a product of Green's function involving intermediate angles that are integrated over. This leads to the final result in the second line of \ref{aol.pexp1}

\end{document}